\documentclass[11pt]{article}
\usepackage{amsfonts,amsmath,amssymb}
\usepackage{enumerate}
\usepackage[hidelinks]{hyperref}
\usepackage{bbm}
\usepackage{nicefrac}
\usepackage[all]{xy}
\usepackage{graphicx}
\usepackage{subcaption}
\usepackage{bm}
\usepackage{makecell}
\usepackage[table]{xcolor}

\usepackage{upgreek}
\usepackage{slashed}

\usepackage{float}			% NEW Helps position tables right here [H]
\usepackage{lscape}                 % NEW creates pages in landscape format

\usepackage{booktabs}

\newcommand{\be}{\begin{equation}}
\newcommand{\ee}{\end{equation}}

\newcommand{\bea}{\begin{eqnarray}}
\newcommand{\eea}{\end{eqnarray}}

\newcommand{\bes}{\begin{subequations}}
\newcommand{\ees}{\end{subequations}}

\newcommand{\cN}{{\cal N}}

\newcommand{\cS}{{\cal S}}
\newcommand{\cA}{{\cal A}}

\def\sst#1{{\scriptscriptstyle #1}}

\def\0{{\sst{(0)}}}
\def\1{{\sst{(1)}}}
\def\2{{\sst{(2)}}}
\def\3{{\sst{(3)}}}
\def\4{{\sst{(4)}}}
\def\5{{\sst{(5)}}}
\def\6{{\sst{(6)}}}
\def\7{{\sst{(7)}}}
\def\8{{\sst{(8)}}}

\def\cV{{{\cal V}}}
\def\cM{{{\cal M}}}
\def\cY{{{\cal Y}}}

\newcommand{\cT}{{\cal T}}

\newcommand{\cL}{{\cal L}}

\newcommand{\vol}{\textrm{vol}}

\allowdisplaybreaks  %Introduces page breaks inside \eqnarray%

\usepackage{multirow}
\usepackage{rotating}

\newcommand{\ba}{\begin{align}}
\newcommand{\ea}{\end{align}}

\newcommand{\bse}{\begin{subequations}}
\newcommand{\ese}{\end{subequations}}

\newcommand{\fT}{\mathcal{T}}
\newcommand{\GLg}[1]{\mathrm{GL}( #1 )}

\newcommand{\Ex}[1]{\mathrm{E}_{#1(#1)}}

\usepackage{lipsum}

\newlength\Colsep
\newcommand{\uA}{{\underline{A}}}
\newcommand{\uB}{{\underline{B}}}
\newcommand{\uC}{{\underline{C}}}
\newcommand{\uD}{{\underline{D}}}
\newcommand{\uE}{{\underline{E}}}
\newcommand{\uF}{{\underline{F}}}
\newcommand{\uG}{{\underline{G}}}

\newcommand{\uK}{{\underline{K}}}
\newcommand{\uL}{{\underline{L}}}
\newcommand{\uM}{\underline{M}}
\newcommand{\uN}{\underline{N}}
\newcommand{\uP}{\underline{P}}
\newcommand{\uQ}{\underline{Q}}
\newcommand{\uR}{\underline{R}}
\newcommand{\uS}{\underline{S}}
\newcommand{\uT}{\underline{T}}

\begin{document}

\addtocontents{toc}{\setcounter{tocdepth}{2}}

\makeatletter
\renewcommand{\theequation}{\thesection.\arabic{equation}}
\@addtoreset{equation}{section}
\makeatother

\begin{titlepage}

\begin{flushright}
%
%IFT-UAM/CSIC-26-xx
%
%\today
\end{flushright}

\vspace{20pt}

   \begin{center}
   \baselineskip=16pt

   \begin{Large}\textbf{
Maximal $D=5$ trombone supergravity from M5-branes
\\[8pt]
and $\mathrm{SU}(2)$-flavoured $\mathcal{N}=1$ class $\mathcal{S}$ operator spectra 
}
   \end{Large}

\vspace{25pt}

{\large Ritabrata Bhattacharya$^{1,2}$, Abhay Katyal$^{2}$ and Oscar Varela$^{2,3}$ }

\vspace{30pt}

	\begin{small}

	{\it $^1$ Department of Physics, Indian Institute of Science Education and Research Bhopal \\
	Bhopal, 462066, India} 

	\vspace{15pt}
          
   {\it $^2$ Department of Physics, Utah State University, Logan, UT 84322, USA} 

	\vspace{15pt}
          
	{\it $^3$  Instituto de F\'\i sica Te\'orica UAM/CSIC, 28049 Madrid, Spain} 
		
	\end{small}

\vskip 70pt

\end{center}

\begin{center}
\textbf{Abstract}
\end{center}

\begin{quote}

We recently presented a new $D=5$ $\mathcal{N}=8$ gauged supergravity involving the local trombone scaling symmetry. It arises by consistent truncation of M-theory on the internal space of the Maldacena-N\'u\~nez AdS$_5$ solution dual to the $\cN=2$ four-dimensional superconformal field theory (SCFT) of class $\mathcal{S}$ associated to M5-branes wrapped on an unpunctured Riemann surface. Using exceptional generalised geometry/field theory, we extend that construction to show that the same $D=5$ $\mathcal{N}=8$ supergravity also arises by consistent truncation of $D=11$ supergravity on the family of $\mathcal{N}=1$ M5-brane-wrapped solutions of Bah-Beem-Bobev-Wecht, including the $\mathcal{N}=1$  Maldacena-N\'u\~nez (MN1) configuration. Then, using recently derived mass matrices, we compute universal sectors of the Kaluza-Klein spectrum on the MN1 solution. In general, this universal spectrum is only locally defined, and we give a prescription for extracting globally defined subsectors thereof. This globally defined universal Kaluza-Klein spectrum is dual to a universal sector of the light operator spectrum of the SU(2)-flavoured $\mathcal{N}=1$ class $\mathcal{S}$ SCFT dual to MN1.

\end{quote}

\vfill

\end{titlepage}

\tableofcontents

%%%%%%%%%%%%%%%%%%%%%%%%%%%%%%%%%%%%%%%%%%%%%%%%%%%%%%%%%%%%%%%%%%%%%%%%%%

%%%%%%%%%%%%%%%
%%%%%%%%%%%%%%%

\section{Introduction}

%%%%%%%%%%%%%%%
%%%%%%%%%%%%%%%

Four-dimensional $\mathcal{N}=1$ superconformal field theories (SCFTs) of class $\mathcal{S}$ and type A$_{N-1}$ arise from twisted compactifications of the six-dimensional $(2,0)$ theory on a stack of $N$ M5-branes wrapped on a Riemann surface $\Sigma_2$ of genus $g$ \cite{Benini:2009mz,Bah:2011vv,Bah:2012dg}. In the unpunctured family considered here, introduced by Bah, Beem, Bobev and Wecht (BBBW) \cite{Bah:2011vv,Bah:2012dg}, the normal bundle to $\Sigma_2$ splits into two line bundles of degrees $p$ and $q$, constrained by $p+q=2g-2$, and the topological twist preserves four supercharges in four dimensions. The resulting family is conveniently labelled by the rational parameter
\begin{equation} \label{eq:defz}
z \equiv \frac{p-q}{p+q} \; .
\end{equation}
At generic $z$, the SCFT has a U$(1)_+$ flavour symmetry in addition to its U$(1)_-$ R-symmetry, while at $z=0$ the flavour symmetry enhances to SU$(2)_+$, with the subscripts $\pm$ motivated in the main text. These theories are generically strongly coupled and need not possess conventional weakly coupled Lagrangian descriptions, but a substantial amount of protected information can nevertheless be inferred from their six-dimensional origin.

At large $N$, this family admits a holographic description in terms of the AdS$_5\times(\Sigma_2\rtimes_{p,q}S^4)$ solutions of $D=11$ supergravity constructed by BBBW \cite{Bah:2011vv,Bah:2012dg}. The symbol $\rtimes_{p,q}$ indicates that the four-sphere is fibred over $\Sigma_2$ according to the two twisting integers $p$ and $q$, and we will focus on the case where $\Sigma_2=H^2/\Gamma$ is a compact hyperbolic Riemann surface, with $\Gamma$ a discrete group of isometries. These backgrounds arise near the horizon of the wrapped M5-branes, with the $S^4$ encoding the transverse directions and its fibration implementing the twist. The generic BBBW solution preserves $\mathcal{N}=1$ supersymmetry and U$(1)_+$ flavour symmetry. Two distinguished limits recover the wrapped-M5 solutions of Maldacena and N\'u\~nez (MN) \cite{Maldacena:2000mw}: the $z=0$, $p=q$ configuration is the $\mathcal{N}=1$ MN1 solution, with enhanced SU$(2)_+$ flavour symmetry, whereas the $z=1$, $q=0$ endpoint is the MN2 solution with augmented $\mathcal{N}=2$ supersymmetry, see also \cite{Gaiotto:2009we,Gaiotto:2009gz}. 

Much of the information presently available about these theories is encoded in coarse protected data, including 't Hooft anomalies \cite{Bah:2020uev} and conformal central charges \cite{Baggio:2014hua}. Further information is encoded in quantities such as the superconformal index \cite{Beem:2012yn} that capture protected sectors of the operator spectrum and have been used to reproduce microscopically the entropy of supersymmetric AdS$_5$ black holes \cite{Bobev:2022ocx,David:2025pby}. Such data provide stringent and often exact characterisations of the SCFTs, but they do not resolve the spectrum operator by operator. In particular, the conformal dimensions and flavour and R-charges of light single-trace operators, whose dimensions remain of order one as $N$ becomes large, are much less well understood. The enhanced-supersymmetry MN2 endpoint is better controlled: we accessed its universal Kaluza-Klein (KK) spectrum in \cite{Bhattacharya:2024tjw,Bhattacharya:2026iaa}, thereby characterising the light operator spectrum of the dual $\cN=2$ class $\cS$ SCFT. Extending comparable operator-level information away from MN2 into the entire BBBW, $\cN=1$ class $\cS$ family is an important problem, since the spectrum of local operators, together with the operator-product expansion coefficients, constitutes the fundamental dynamical data of any SCFT.

This paper has two principal results. First, we construct a local maximally supersymmetric consistent truncation of $D=11$ supergravity on every unpunctured BBBW geometry $\Sigma_2\rtimes_{p,q}S^4$. All members of the family reduce to the same $D=5$ $\mathcal{N}=8$ TCSO$(5,0,1;1)$-gauged supergravity of \cite{Bhattacharya:2024tjw,Varela:2025xeb}, with the twisting integers entering through the duality frame selected by the topological twist. A distinctive feature of this maximal five-dimensional supergravity \cite{Bhattacharya:2024tjw,Varela:2025xeb} is that it involves a gauging of the trombone scaling symmetry. Second, specialising this construction to the $z=0$ MN1 vacuum, and employing the trombone-enhanced KK mass matrices derived in \cite{Pico:2026kch} following \cite{Malek:2019eaz,Malek:2020yue,Varela:2020wty,Cesaro:2020soq}, we determine a globally defined, $\Sigma_2$-constant, universal sector of the KK spectrum over MN1 at arbitrary level. The corresponding operators form infinite towers of $\mathrm{SU}(2,2|1)\times\mathrm{SU}(2)_+$ graviton, gravitino and vector multiplets whose dimensions are governed by a single closed expression and whose multiplicities are determined exactly at all KK levels. By a basic entry in the AdS/CFT dictionary \cite{Maldacena:1997re,Gubser:1998bc,Witten:1998qj}, our second result characterises holographically a sector of the light single-trace operator spectrum of the MN1 SCFT. The KK spectral analysis of the generic BBBW vacua is presented separately in \cite{Bhattacharya:2026iaa}.

We will do both things within the framework of exceptional field theory/generalised geometry (ExFT/ExGG) \cite{Hohm:2013pua,Hohm:2013vpa,Coimbra:2011ky,Coimbra:2012af}. These formulations make manifest the exceptional symmetry E$_{d(d)}$ of maximal supergravity and combine the higher-dimensional diffeomorphism and form-field gauge symmetries into generalised geometric objects. When the AdS background belongs to a consistent truncation to a maximally supersymmetric $D$-dimensional supergravity, the linearised fields of that theory form the lowest level in the KK expansion. Higher KK levels can then be generated by expanding all ExFT fluctuations in a single basis of scalar functions, equivalently in eigenfunctions of the spin-2 operator, rather than in separate harmonic bases for every higher-dimensional field. The representation matrices governing the action of the generalised frame on this basis, together with the lower-dimensional embedding tensor and vacuum scalar matrix, determine algebraic mass matrices for all spins. The original differential spectral problem is thereby converted into the diagonalisation of universal matrices that are infinite-dimensional in total but block-diagonal into finite-dimensional sectors at each KK level \cite{Malek:2019eaz,Malek:2020yue,Varela:2020wty,Cesaro:2020soq}.

The BBBW backgrounds \cite{Bah:2011vv,Bah:2012dg} in general, and MN1 in particular \cite{Maldacena:2000mw}, become amenable to this strategy through the maximal five-dimensional supergravity and its eleven-\-di\-men\-sion\-al origin developed here. There is, however, an important global subtlety. The generalised identity structure underlying our maximal truncation is constructed locally by replacing the compact surface $\Sigma_2=H^2/\Gamma$ with the locally equivalent non-compact group manifold $B_2$ of the non-abelian two-dimensional Lie algebra. Its non-unimodularity induces a gauging of the trombone scaling symmetry in the five-dimensional TCSO$(5,0,1;1)$ theory and implies that the KK states obtained directly from the maximal ExFT construction need not extend globally over $\Sigma_2\rtimes_{p,q}S^4$. We therefore regard this local spectrum as a putative reservoir of candidate modes rather than, automatically, as the physical spectrum. A globally defined generalised U$(1)_z$-structure is nevertheless available on every BBBW bundle \cite{Cassani:2019vcl,Cassani:2020cod}, see also \cite{Josse:2025uro}. At the MN1 point, requiring invariance under U$(1)_0$ selects a bona fide global, $\Sigma_2$-constant subsector of the putative spectrum, which also organises itself into  $\mathrm{SU}(2,2|1)\times\mathrm{SU}(2)_+$ representations. A direct analysis of the eleven-dimensional graviton equation further shows how additional global modes, with non-trivial dependence on $\Sigma_2$, are governed by weighted Maass operators.

The paper is organised as follows. Section~\ref{sec:4DSugra} reviews $D=5$ $\cN=8$ TCSO$(5,0,1;1)$-gauged supergravity and constructs its new $D=11$ origin by consistent truncation on the $\Sigma_2\rtimes_{p,q}S^4$ twisted manifolds of BBBW. Section~\ref{sec:Subsectors} reviews some relevant local and global generalised structures on $\Sigma_2\rtimes_{p,q}S^4$, and uses this language to recover some previously known submaximal consistent truncations on BBBW as subsectors of our construction. In section~\ref{sec:Spectra}, we explain the distinction between putative and global modes and determine the all-level U$(1)_0$-invariant, global KK spectrum on MN1. Section~\ref{sec:Discussion} concludes. Further conventions, the trombone mass matrices, the local putative MN1 spectrum and a direct eleven-dimensional analysis of the graviton tower are collected in the appendices.

%%%%%%%%%%%%%%%
%%%%%%%%%%%%%%%

\section{$\mathrm{TCSO}(5,0,1;1)$ supergravity and its $D=11$ origin} \label{sec:4DSugra}

%%%%%%%%%%%%%%%
%%%%%%%%%%%%%%%

Let us start by reviewing some relevant aspects of the TCSO$(5,0,1;1)$ gauging of $D=5$ $\cN=8$ supergravity, before discussing the $D=11$ origin of this theory. As an intermediate step, we first show that $D=5$ $\cN=8$ TCSO$(5,0,1;1)$-gauged supergravity arises by consistent truncation of $D=11$ supergravity on the direct product manifold $B_2 \times S^4$. Building on this auxiliary result, we finally construct the truncation on the BBBW twisted manifold $\Sigma_2 \rtimes_{p,q} S^4$.

%%%%%%%%%%%%%%%

\subsection{$D=5$ $\cN=8$ $\mathrm{TCSO}(5,0,1;1)$-gauged supergravity} \label{sec:FieldContentET}

%%%%%%%%%%%%%%%

Recall that the bosonic and fermionic fields of maximal supergravity in five spacetime dimensions come in representations of $\mathbb{R}^+ \times \mathrm{E}_{6(6)}$ and $\mathbb{R}^+ \times \mathrm{USp}(8)$, respectively. Here, $\mathbb{R}^+$ is the trombone scaling symmetry, $\mathrm{E}_{6(6)}$ is the duality symmetry group of the ungauged theory \cite{Cremmer:1979uq}, and $\mathrm{USp}(8)$ its maximal compact subgroup. The metric, $g_{\mu \nu}$, gauge fields, $A_\mu{}^{\uM}$, and two-form potentials, $B_{\mu\nu \, \uM}$, respectively sit in the $\bm{1}_2$, $\overline{\bm{27}}_1$ and $\bm{27}_2$ of $\mathbb{R}^+ \times \mathrm{E}_{6(6)}$, while the $\bm{42}_0$ scalars parametrise a coset representative, $\cV_{\uM}{}^{\uN}$, of $\mathrm{E}_{6(6)} / \mathrm{USp}(8)$, with metric $M_{\uM \uN} = (\cV \cV^{\mathrm{T}})$. The gravitini, $\psi_\mu{}^i$, and spin-$1/2$ fermions transform in the $\overline{\bm{8}}_{\frac{1}{2}}$ and $\overline{\bm{48}}_{-\frac{1}{2}}$ of $\mathbb{R}^+ \times \mathrm{USp}(8)$. Indices $\mu=0, \ldots, 4$, often suppressed, and $\uM= 1 , \ldots , 27$, $i = 1, \ldots , 8$ respectively label spacetime vectors, the fundamental of $\mathrm{E}_{6(6)}$ and the fundamental of $\mathrm{USp}(8)$. Five-dimensional spinor indices are always suppressed. In any case, we will mostly focus on the bosons in the following.

Only in ungauged supergravity are $\mathbb{R}^+ \times \mathrm{E}_{6(6)}$ and $\mathbb{R}^+ \times \mathrm{USp}(8)$ (global) symmetries. In the gauged theory, the symmetry is reduced to a locally-realised subgroup of $\mathbb{R}^+ \times \mathrm{E}_{6(6)}$. In this paper, we will focus on $D=5$ $\cN=8$ supergravity equipped with the seventeen-dimensional gauge group \cite{Varela:2025xeb}
\begin{equation} \label{eq:ProdTCSO}
\textrm{TCSO} (5,0,1;1) \equiv B_2 \ltimes \textrm{ISO} (5) \equiv \big( B_2 \times \textrm{SO} (5) \big) \ltimes \mathbb{R}^{5}  \,  \subset \, \mathbb{R}^+ \times \mathrm{E}_{6(6)} \; .
\end{equation}
Here, $B_2$ is the Borel subgroup of SL$(2, \mathbb{R})$, namely, (the group corresponding to) the two-dimensional non-abelian Lie algebra, and $\textrm{ISO}(5) \equiv \textrm{CSO} (5,0,0) \equiv \textrm{SO} (5)  \ltimes \mathbb{R}^{5}$ is the usual Euclidean group in five dimensions. We refer to \cite{Varela:2025xeb} for details on the semidirect actions specified in (\ref{eq:ProdTCSO}) by the symbol $\ltimes$. 

While a formulation of this theory exists with only 17 gauge fields in the adjoint of $\textrm{TCSO} (5,0,1;1)$ along with 10 two-forms, see \cite{Varela:2025xeb}, we instead choose to work in the embedding tensor formalism \cite{deWit:2004nw,LeDiffon:2008sh,Varela:2025xeb}. We  thereby maintain formal $\mathbb{R}^+ \times \mathrm{E}_{6(6)}$ covariance and keep all $\overline{\bm{27}}_1$ gauge fields and $\bm{27}_2$ two-forms. In this approach, all local gauging-related interactions are introduced by the embedding tensor. The latter will be denoted either by $X_{\uM}$ or by $X_{\uM \uN}{}^{\uP}$, either omitting or displaying its E$_{6(6)}$ representation indices, depending on convenience. The embedding tensor is a constant object in the $\bm{27}_{-1} + \overline{\bm{351}}_{-1}$ of $\mathbb{R}^+ \times \mathrm{E}_{6(6)}$, which takes values in the Lie algebra of $\mathbb{R}^+ \times \mathrm{E}_{6(6)}$, and governs how the  Lie algebra of the gauge group (\ref{eq:ProdTCSO}) is embedded in the former. It is also subject to the quadratic constraint
\begin{equation}
[X_{\uM} , X_{\uN} ] = -X_{\uM \uN}{}^{\uP} \, X_{\uP} \; , 
\end{equation}
which ensures closure of the gauge algebra. Finally, only equivalence classes of embedding tensors are physically distinct. Namely, the same gauged supergravity can be described by either of the embedding tensors, $X_{\uM \uN}{}^{\uP}$ or $\tilde{X}_{\uM \uN}{}^{\uP}$, related by an $\mathrm{E}_{6(6)}$ transformation $U_{\uM}{}^{\uN}$ via
\begin{equation} \label{eq:TransXSymbol}
\tilde{X}_{\uM \uN}{}^{\uP} = U_{\uM}{}^{\uQ} \, U_{\uN}{}^{\uR} \, X_{\uQ \uR}{}^{\uS}  \, (U^{-1})_{\uS}{}^{\uP} \; .
\end{equation}

A choice of embedding tensor within the same duality class (\ref{eq:TransXSymbol}) defines a so-called duality frame. Regardless of the duality frame employed to write a $D=5$ $\cN=8$ gauged supergravity, the latter is completely specified by its embedding tensor. We will find it helpful to use various duality frames to describe the same $D=5$ $\cN=8$ TCSO$(5,0,1;1)$-gauged supergravity. In the defining frame of \cite{Varela:2025xeb}, the embedding tensor splits as $X_{\uM} = (X_{AB} , X^{xA})$, with indices $A=1 , \ldots, 6$ and $x=1,2$ in the fundamental of SL$(6 , \mathbb{R})$ and SL$(2 , \mathbb{R})$, respectively, and
\begin{equation} \label{eq:XSymbols}
X_{AB} = 2 \, \theta_{C[A} \, t_{B]}{}^C \; , \qquad
X^{xA} = \tfrac{8}{3} \, \xi^{xA} \, \, \mathbf{1} +8 \,  \xi^{xB} \, t_B{}^A - 8\,  \xi^{yA} \, t_y{}^x \; .
\end{equation}
Here, $ \mathbf{1}$ and $t_\alpha = (t_A{}^B , t_x{}^y ,  t_{xABC} )$ are the generators of the trombone $\mathbb{R}^+$ and of E$_{6(6)}$, respectively, with representation indices omitted for legibility, in the conventions of appendix A of \cite{Varela:2025xeb}. Finally, the only non-vanishing components of $\theta_{AB} = \theta_{(AB)}$ and $\xi^{xA}$ are along the SO(5)-invariant metric and the $B_2$ structure constants, respectively, such that
\begin{equation} \label{eq:Charges1}
\theta_{AB} = \big( g_1 \, \delta_{ij}  \, , \, \theta_{i6} = 0  \, , \, \theta_{66} = 0   \big) \; , \qquad 
\xi^{xA} =
\left(
\begin{array}{ll}
\xi^{1i} = 0 &  \xi^{16} = -\tfrac18 \, g_2  \\
\xi^{2i} = 0 & \xi^{26} =0
 \end{array} \right) \; ,
\end{equation}
with $i=1, \ldots , 5$ (recycled from the fermion index above), and $g_1$, $g_2$ non-vanishing coupling constants. 

We will also find it useful to describe $D=5$ $\cN=8$ TCSO$(5,0,1;1)$-gauged supergravity using a more general family of duality frames in which the embedding tensor depends on two integers, $p$,  $q$. In these frames, the embedding tensor still splits as $\tilde{X}_{\uM} = (\tilde{X}_{AB} , \tilde{X}^{xA})$, where now \cite{Varela:2025xeb}
{\setlength\arraycolsep{2pt}
\begin{eqnarray} \label{eq:XSymbolsTilde}
\tilde{X}_{AB} & =&  2 \, \theta_{C[A} \, t_{B]}{}^C + \theta_{AB}{}^C{}_D \, t_C{}^D -2 \,  \xi^{xCD}{}_{[A|} \, t_{x|B] CD}   \; ,  \\[5pt]
\tilde{X}^{xA}  & =&  \tfrac{8}{3} \, \xi^{xA} \, \, \mathbf{1} +8 \,  \xi^{xB} \, t_B{}^A - 8\,  \xi^{yA} \, t_y{}^x  +2 \,  \xi^{xAB}{}_C \, t_B{}^C   -\tfrac{1}{36} \,  \epsilon^{xy} \,  \epsilon^{BCDEFG} \, \theta_{EF}{}^A{}_G \, t_{y BCD} \; . \nonumber
\end{eqnarray}
}Here, $\theta_{AB}$ and $\xi^{xA}$ are still given by (\ref{eq:Charges1}), while $\theta_{AB}{}^C{}_D = \theta_{[AB}{}^C{}_{D]}$, $\theta_{AB}{}^C{}_C =0$ and $\xi^{xAB}{}_C =\xi^{x[AB]}{}_C$, $\xi^{xAB}{}_B=0$ have non-vanishing components
\begin{eqnarray} \label{eq:ExtraCompsET}
\theta_{12}{}^6{}_5 = 6  \tfrac{q}{p +q} g_2^2 g_1^{-1} \;  , \quad
\theta_{34}{}^6{}_5 = -6 \tfrac{p}{p +q} g_2^2 g_1^{-1} \; , \quad
\xi^{212}{}_{6} = -\tfrac{p}{p +q} g_2  \;  , \quad
\xi^{234}{}_{6} =  \tfrac{q}{p +q} g_2 \;  . 
\end{eqnarray}
up to antisymmetric permutations. The frames (\ref{eq:XSymbolsTilde}) and (\ref{eq:XSymbols}) are related through a duality transformation 
(\ref{eq:TransXSymbol}) with E$_{6(6)}$ element \cite{Varela:2025xeb}
\begin{equation} \label{eq:E7GroupElement}
U_{\uM}{}^{\uN} =e^{\Upsilon_{\uM}{}^{\uN}} \; , \quad \textrm{with} \quad 
\Upsilon = g_2 \, g_1^{-1} \, \tfrac{1}{p+ q}  \big( p \,  t_{2126} -  q \, t_{2346} \big) \; .
\end{equation}

The embedding tensor (\ref{eq:XSymbols}) can be regarded as the $p=q=0$ member of the family (\ref{eq:XSymbolsTilde}). The limit $p=q=0$ at fixed, finite $p+q$, sets the duality transformation (\ref{eq:E7GroupElement}) to $\Upsilon =0$, $U =  \mathbf{1}$, and (\ref{eq:XSymbolsTilde}) reduces to (\ref{eq:XSymbols}) as required by (\ref{eq:TransXSymbol}). Indeed, in this limit, $\theta_{AB}{}^C{}_D=0$, $\xi^{xAB}{}_C=0$, as can be seen from (\ref{eq:ExtraCompsET}). Let us reiterate that both embedding tensors (\ref{eq:XSymbols}) and (\ref{eq:XSymbolsTilde}) describe the same $D=5$ $\cN=8$ supergravity. Either form may be more convenient to describe the $D=11$ origin of this theory by reduction on different six-dimensional internal manifolds. In \cite{Bhattacharya:2024tjw}, we showed that the TCSO$(5,0,1;1)$ theory arises by consistent truncation on the internal manifold corresponding to MN2. In section~\ref{sec:TruncTwisted} we will extend that result to the entire BBBW family.

%%%%%%%%%%%%%%%%

\subsection{Consistent $\cN=8$ truncation of $D=11$ supergravity on $B_2 \times S^4$} 

\label{sec:TruncDP}

%%%%%%%%%%%%%%%%

We will use the language of ExGG/ExFT \cite{Hohm:2013pua,Hohm:2013vpa,Coimbra:2011ky,Coimbra:2012af} to characterise maximally supersymmetric consistent truncations. We refer the reader to appendix~\ref{sec:Frames_EGGintro} for a brief review of the E$_{6(6)}$ ExGG reformulation of $D=11$ supergravity. In this framework, a consistent truncation of $D=11$ supergravity on a six-dimensional manifold $M_6$ to a $D=5$ supergravity exists if the generalised  USp$(8)$-structure specified by the generalised metric is reduced to a local generalised $G$-structure whose intrinsic torsion is a constant, E$_{6(6)}$ singlet \cite{Lee:2014mla,Cassani:2019vcl}. In particular, a constant-torsion generalised identity structure (GIS), $G=\bm{1}$, gives rise to a maximally supersymmetric consistent truncation \cite{Lee:2014mla}. Thus, a consistent truncation of $D=11$ supergravity on a six-dimensional manifold $M_6$ to $D=5$ $\cN=8$ supergravity exists if the ExGG fields $g_{\mu \nu}(x,y)$, $(G^{-1})^{MN} (x,y)$, $\cA_\mu{}^M (x,y)$, etc.,  in (\ref{eq:11DFields}),  \eqref{eq:ExObjects} factorise in terms of the five-dimensional fields $g_{\mu\nu}(x)$, $M^{\uM\uN}(x)$, $A^{\uM}(x)$, etc., reviewed in section~\ref{sec:FieldContentET}, via the generalised Scherk--Schwarz ansatz \cite{Berman:2012uy,Lee:2014mla}
\begin{eqnarray} \label{eq:GenSS}
& \tilde \Delta(x,y) = \tilde \Delta(y) \; , \qquad
g_{\mu \nu}(x,y)=  g_{\mu \nu}(x) \; , \nonumber \\[4pt]
& (G^{-1})^{MN} (x,y) =M^{\uP\uQ}(x) \, \hat{E}^M{}_{\uP}(y) \hat{E}^N{}_{\uQ}(y) \; , \quad 
\cA_\mu{}^M (x,y) = \hat{E}^M{}_{\uN}(y) \, A_\mu{}^{\uN}(x) \; , \;
\end{eqnarray}
and (\ref{eq:IntTor}) holds with constant intrinsic torsion $X_{\uM\uN}{}^{\uP}$  \cite{Lee:2014mla,Cassani:2019vcl}, namely,
\begin{equation} \label{eq:CTConst}
L_{\hat E_{\uM} (y)} \hat E_{\uN}(y) = - X_{\uM\uN}{}^{\uP} \, \hat E_{\uP} (y)  \; .
\end{equation}
The constant intrinsic torsion $X_{\uM\uN}{}^{\uP}$ then defines the embedding tensor of the $D=5$ $\cN=8$ supergravity \cite{Berman:2012uy,Lee:2014mla}. In (\ref{eq:GenSS}), (\ref{eq:CTConst}), $x$, $y$ label five-dimensional external and $M_6$ internal coordinates, respectively, and $\hat{E}^M{}_{\uN}(y)$ is the generalised vielbein on $M_6$ defined in general by (\ref{eq:GenMet}).

Equipped with these prescriptions, we will show in section~\ref{sec:TruncTwisted} that $D=11$ supergravity admits a maximally supersymmetric consistent truncation on the twisted internal manifold $\Sigma_2 \rtimes_{p,q} S^4$ associated with the supersymmetric AdS$_5$ solutions of BBBW \cite{Bah:2011vv,Bah:2012dg}. In the present section, we begin by establishing the existence of a related $D=11$ truncation on the direct-product manifold $M_6 = B_2 \times S^4$, where $B_2$ is the group manifold of the non-abelian two-dimensional Lie algebra. This truncation gives rise to $D=5$ $\cN=8$ TCSO$(5,0,1;1)$-gauged supergravity. By the discussion above, showing the existence of this truncation amounts to constructing a constant-intrinsic-torsion GIS on $B_2 \times S^4$. Equivalently, we must build an inverse generalised frame $\hat{E}^{M}{}_{\uN}(y)$ from geometric data on $B_2 \times S^4$, and then verify that \eqref{eq:CTConst} holds, with $X_{\uM}$ given by the TCSO$(5,0,1;1)$ embedding tensor. The basic ingredients of the GIS on $B_2 \times S^4$ are the conventional parallelisation of the group manifold $B_2$ and the generalised parallelisation of $S^4$ constructed in \cite{Lee:2014mla}, as we will momentarily see.

In order to build this GIS, it is useful to regard the inverse generalised frame $\hat{E}^{M}{}_{\uN}(y)$ as a set of $\bm{27}$ generalised vectors, labelled by the flat index $\uN$, each carrying a curved generalised-vector index $M$. For fixed $\uN$, each such generalised vector decomposes, according to \eqref{eq:27Branching}, into GL$(6)$-covariant components as in the rightmost relation of \eqref{eq:ExObjects},
\begin{equation} \label{eq:GenFrameSplit}
\hat{E}^M{}_{\uN}  = \big( \hat{E}^m{}_{\uN}  , \hat{E}_{mn \, \uN}  ,  \hat{E}_{mnpqr \, \uN}  \big) \; .
\end{equation}
To define each of these components, it is helpful to also split the flat index $\uN$ under, this time, $\mathbb{R}^+ \times  \textrm{E}_{6(6)} \supset \textrm{GL}(2) \times \textrm{SL}(5)$ instead, because the GL(6)-covariant quantities on $B_2 \times S^4$ that we will use come in representations of that group. Under $\mathbb{R}^+ \times  \textrm{E}_{6(6)} \supset \textrm{GL}(2) \times \textrm{SL}(5)$, and omitting $\mathbb{R}^+$ charges,
\begin{equation} \label{eq:SplitGL3SL5}
\bm{27} \rightarrow (\bm{2} , \bm{1}) + (\bm{2} , \bm{5}^\prime) + (\bm{1} , \bm{5}) + (\bm{1} , \bm{10}) \; ,
\end{equation}
so that, at fixed value of the curved index $M$, the flat index $\uN$ accordingly branches as
\begin{equation} \label{eq:GenFrameSplit2}
\hat{E}^M{}_{\uN}  = \big( \hat{E}^M{}_x , \, \hat{E}^{M x i}  , \, \hat{E}^M{}_i , \,  \hat{E}^M{}_{ij} \big) \; ,
\end{equation}
with $x=1,2$, $i = 1, \ldots , 5$ as in section~\ref{sec:FieldContentET}. Finally, we take the following expressions for the various blocks in (\ref{eq:GenFrameSplit2}):
{\setlength\arraycolsep{2pt}
\begin{eqnarray} \label{eq:GenFrameB2S4}
&& \hat{E}_x = ( \hat{e}_x ,\;  0  , \; 0) \; , \nonumber \\[5pt]
&& \hat{E}^{ x i}   =   ( 0, \; R\, e^x \wedge d y^i,  \; e^x \wedge (- y^i \textrm{vol}_{4} + R\, d y^i \wedge A ))   ,  \nonumber \\[5pt]
&& \hat{E}_i =
\left(
0,\;
y_i \textrm{vol}_2,\;
\textrm{vol}_2 \wedge (R*_4 d y_i + y_i A)
\right)   \; ,  \nonumber \\[5pt]
&& \hat{E}_{ij} =  ( v_{ij} , \; R^2*_4\! (d y_i\wedge d y_j) + \iota_{v_{ij}} A  , \;  0)  \; ,
\end{eqnarray}
}with the curved index $M$ omitted on the l.h.s.'s, and split on the r.h.s.'s as in (\ref{eq:GenFrameSplit}). Here, $e^x$ is a vielbein one-form with inverse vector field $\hat{e}_x$, and $\textrm{vol}_2$ a volume form, all of them defined on $B_2$. In a convenient set of coordinates $(x,y)$ (where we are recycling these symbols w.r.t.~the discussion below (\ref{eq:CTConst})) these, together with the associated volume form $\textrm{vol}_2$ and metric $g_\Sigma$ on $B_2$, can be taken to be
\begin{equation} \label{eq:B2Geom}
e^1 = R_\Sigma \,  \frac{dx}{y} \; , \qquad 
e^2 = R_\Sigma \, \frac{dy}{y} \; , \qquad 
\textrm{vol}_2 = R_\Sigma^2 \, \frac{dx \wedge dy}{y^2} \; , \qquad 
g_\Sigma = R_\Sigma^2 \, \frac{dx^2 +dy^2}{y^2} \; .
\end{equation}
The $S^4$ quantities appearing in (\ref{eq:GenFrameB2S4}) include the coordinates $y^i$, constrained as $\delta_{ij} y^i y^j = 1$, the SO(5) Killing vectors $v^{ij}$, the volume form $\textrm{vol}_4$, and a local three-form potential $A$ for the latter, $dA = 3 R^{-1} \, \vol_4$. The Hodge dual $*_4$ is taken w.r.t.~the round metric on $S^4$, and $\imath_{v}$ denotes the interior product w.r.t.~a conventional vector $v$. The constant $R$ is the $S^4$ radius and $R_\Sigma$ in (\ref{eq:B2Geom}) sets a scale on $B_2$.

As anticipated above, the generalised frame \eqref{eq:GenFrameSplit2}, \eqref{eq:GenFrameB2S4} is built from the ordinary parallelisation of $B_2$ \cite{Scherk:1979zr} and the generalised parallelisation of $S^4$ \cite{Lee:2014mla}. It also makes use of various generalised tensors on $S^4$ introduced in \cite{Lee:2014mla,Cassani:2019vcl}. Taking the generalised Lie derivative \eqref{eq:LieDer} of the generalised frame \eqref{eq:GenFrameSplit2}, \eqref{eq:GenFrameB2S4} w.r.t.~itself, a lengthy calculation shows that this frame indeed satisfies the GIS condition \eqref{eq:CTConst}. The coordinate form of the frame, its complete generalised Lie derivative algebra and the explicit comparison with the embedding tensor are collected in appendix~\ref{sec:GeomGISAppendix}. The corresponding constant intrinsic torsion $X_{\uM\uN}{}^{\uP}$ is precisely the TCSO$(5,0,1;1)$ embedding tensor in the duality frame \eqref{eq:XSymbols}, provided the coupling constants of the $D=5$ supergravity and radii of the reduction geometry are identified as
\begin{equation} \label{eq:CouplingsRadii}
g_1=R^{-1}\;,\qquad
g_2=R_{\Sigma}^{-1}\;.
\end{equation}
This establishes the consistency of the truncation of $D=11$ supergravity on $B_2 \times S^4$ down to $D=5$ $\cN=8$ TCSO$(5,0,1;1)$-gauged supergravity. The explicit embedding of the $D=5$ $\cN=8$ fields into the $D=11$ fields of the conventional formulation of \cite{Cremmer:1978km} can be obtained by bringing the frame (\ref{eq:GenFrameSplit2}),  (\ref{eq:GenFrameB2S4}) to the generalised Scherk-Schwarz expressions (\ref{eq:GenSS}), and then unpacking the ExGG fields using (\ref{eq:ExObjects}). We will give explicit examples in section~\ref{sec:Subsectors}.

%%%%%%%%%%%%%%%%

\subsection{Consistent $\cN=8$ truncation of $D=11$ supergravity on $\Sigma_2 \rtimes_{p,q} S^4$} 

\label{sec:TruncTwisted}

%%%%%%%%%%%%%%%%

Let us now build on the results of section~\ref{sec:TruncDP} and on \cite{Cassani:2019vcl,Cassani:2020cod} to show that $D=5$ $\cN=8$ TCSO$(5,0,1;1)$-gauged supergravity also arises locally by consistent truncation of $D=11$ supergravity on any representative of the family of BBBW twisted six-dimensional manifolds $\Sigma_2 \rtimes_{p,q} S^4$, with $\Sigma_2$ an unpunctured, hyperbolic, constant curvature, genus $g = \frac12 (p+q) +1 \geq 2$ Riemann surface.

Recall that the near-horizon M5-brane solutions $\textrm{AdS}_5 \times (\Sigma_2 \rtimes_{p,q} S^4)$ of BBBW \cite{Bah:2011vv,Bah:2012dg} implement supersymmetry via topological twist \cite{Witten:1988ze,Bershadsky:1995qy}. The fibration is governed by two winding numbers $p$ and $q$, which measure how the U$(1)^2$ Cartan subgroup of the SO(5) isometry of $S^4$ twist over the genus $g$ Riemann surface with the help of the spin connection $\upsilon$ on $\Sigma_2$. The latter is defined in terms of the vielbein $e^x$ on $\Sigma_2$, as usual, by
\begin{equation} \label{eq:SpinConDef}
d e^x
=
-\epsilon^x{}_y\,\upsilon\wedge e^y
\;,\qquad
d\upsilon
=
-\tfrac{1}{R_{\Sigma}^2}\textrm{vol}_2
\; .
\end{equation}
Below we will take these U(1)'s to be generated by the $S^4$ Killing vectors $v^{12}$ and $v^{34}$. We will also trade $\Sigma_2$ locally with the group manifold $B_2$ of the non-abelian two-dimensional Lie algebra. This is a valid step because the two manifolds are locally diffeomorphic, with the relevant vielbein, volume form and metric on $\Sigma_2$ also given locally by (\ref{eq:B2Geom}). In the set of coordinates employed in the latter equation, the spin connection reads, simply,
\begin{equation} \label{eq:SpinCon}
\upsilon = -\frac{dx}{y} \; .
\end{equation}

Now, to show the consistency of the maximally supersymmetric truncation of $D=11$ supergravity on $\Sigma_2 \rtimes_{p,q} S^4$ or, equivalently, $B_2 \rtimes_{p,q} S^4$, we must prove the existence of a family of GISs $\hat{\tilde{E}}^M{}_{\uP} (y)$ on the twisted manifolds $B_2 \rtimes_{p,q} S^4$ with constant intrinsic torsion. In order to write a candidate $\hat{\tilde{E}}^M{}_{\uP} (y)$, we will simply import the prescription of \cite{Cassani:2019vcl,Cassani:2020cod} for the ExGG implementation of the topological twist from the untwisted frame and write
\begin{equation} \label{eq:TwistedFrame}
\hat{\tilde{E}}^M{}_{\uP} (y) = U_N{}^M (y) \, \hat{E}^N{}_{\uP} (y) \; .
\end{equation}
Here, $\hat{E}^N{}_{\uM} (y)$ is the inverse generalised vielbein \eqref{eq:GenFrameSplit2}, \eqref{eq:GenFrameB2S4} on the direct product manifold $B_2 \times S^4$, and $U_N{}^M (y)$ is the local, $B_2 \times S^4$-dependent $\mathbb{R}^+ \times \mathrm{E}_{6(6)}$ transformation \cite{Cassani:2020cod} 
\begin{equation} \label{eq:E6GroupElement}
U_{M}{}^{N} (y) =e^{\Upsilon_{M}{}^{N}(y)} \; , \quad \textrm{with} \quad 
\Upsilon_M{}^N= -\tfrac{R}{p+q} \big(  \tilde{\upsilon} \times_\text{ad} ( p \hat{E}_{12} - q \hat{E}_{34} ) \big)_M{}^N  \; .
\end{equation}
The local $\mathbb{R}^+ \times \mathrm{E}_{6(6)}$ (in fact, $\mathrm{E}_{6(6)}$) Lie algebra element $\Upsilon_M{}^N$ in (\ref{eq:E6GroupElement}) depends on the generalised one-form $\tilde{\upsilon}_M$ and generalised vectors $\hat{E}_{12}$, $\hat{E}_{34}$. The former is the straightforward promotion of the $B_2$ spin connection $\upsilon$ in (\ref{eq:SpinCon}) to ExGG through (the conjugate of) (\ref{eq:27Branching}), namely, $\tilde{\upsilon}_M = (\upsilon ,0 ,0)$. The latter are simply the 12, 34 directions of the components $\hat{E}_{ij}$ in (\ref{eq:GenFrameB2S4}) of the inverse generalised vielbein $\hat{E}^N{}_{\uM} (y)$ on $B_2 \times S^4$. Finally, the operation $\times_\text{ad}$, defined in \cite{Coimbra:2011ky}, ensures that the product of $\tilde{\upsilon}_M$ and $( p \hat{E}^M{}_{12} - q \hat{E}^M{}_{34} )$ is projected onto the Lie algebra of $\mathbb{R}^+ \times \mathrm{E}_{6(6)}$.

Some calculation shows that the GIS $\hat{\tilde{E}}^M{}_{\uN} (y)$ defined in (\ref{eq:TwistedFrame}) also has, like $\hat{E}^M{}_{\uN} (y)$,   constant intrinsic torsion. More concretely, as shown in appendix~\ref{sec:GeomGISAppendix}, it obeys 
\begin{equation} \label{eq:CTConstTilde}
L_{\hat{\tilde{E}}_{\uM} (y)} \hat{\tilde{E}}_{\uN}(y) = - \tilde{X}_{\uM\uN}{}^{\uP} \, \hat{\tilde{E}}_{\uP} (y)  \; .
\end{equation}
Here, $\tilde{X}_{\uM\uN}{}^{\uP}$ is the embedding tensor of the TCSO$(5,0,1;1)$ supergravity in the duality frame specified by (\ref{eq:XSymbolsTilde}) with (\ref{eq:Charges1}), (\ref{eq:ExtraCompsET}). Thus, $D=11$ supergravity admits a local consistent truncation to $D=5$ $\cN=8$ TCSO$(5,0,1;1)$-gauged supergravity on the internal six-dimensional geometries $\Sigma_2 \rtimes_{p,q} S^4$ of the supersymmetric wrapped M5-brane solutions of BBBW \cite{Bah:2011vv,Bah:2012dg}. The MN2 solution \cite{Maldacena:2000mw} is recovered as the $q=0$ instance of BBBW, and our $\cN=8$ truncation reproduces that of \cite{Bhattacharya:2024tjw}. More generally, our construction is valid for all members of the BBBW class.

The fact that $D=11$ supergravity truncates consistently on either direct, $B_2 \times S^4$, or twisted, $B_2 \rtimes_{p,q} S^4$, manifolds, and moreover does so to one and the same five-dimensional supergravity may come as a slight surprise. Of course this is because, as reviewed in section \ref{sec:FieldContentET}, the intrinsic torsions $X_{\uM\uN}{}^{\uP}$ and $\tilde{X}_{\uM\uN}{}^{\uP}$ of the GISs $\hat{E}_{\uP} (y)$, \eqref{eq:GenFrameSplit2}, \eqref{eq:GenFrameB2S4}, on $B_2 \times S^4$ and $\hat{\tilde{E}}_{\uP} (y)$, (\ref{eq:TwistedFrame}), on  $\Sigma_2 \rtimes_{p,q} S^4$ are related by a duality transformation (\ref{eq:TransXSymbol}). Remarkably, this observation says that the local ExGG implementation of the topological twist of \cite{Cassani:2019vcl,Cassani:2020cod} is equivalent, at the level of $D=5$ $\cN=8$ supergravity, to a global duality transformation. Indeed, some calculation leads to the identity
\begin{equation} \label{eq:FlatteningCon}
\Upsilon_{\uM}{}^{\uN} = \Upsilon_{M}{}^{N}(y) \, \hat{E}^M{}_{\uM} (y) \, E_N{}^{\uN} (y) \; , 
\end{equation}
between the constant, $\Upsilon_{\uM}{}^{\uN} $ in (\ref{eq:E7GroupElement}), and the $B_2 \times S^4$-dependent,  $\Upsilon_{M}{}^{N}(y)$ in (\ref{eq:E6GroupElement}), E$_{6(6)}$ Lie algebra elements that respectively implement the duality transformation (\ref{eq:TransXSymbol}) in $D=5$ $\cN=8$ supergravity and the topological twist (\ref{eq:E6GroupElement}) in ExGG. In other words, the former is the flattened version of the latter with the generalised frame $\hat{E}^M{}_{\uM} (y)$, \eqref{eq:GenFrameSplit2}, \eqref{eq:GenFrameB2S4}, on $B_2 \times S^4$. A similar observation about the ExGG prescription \cite{Cassani:2019vcl,Cassani:2020cod} for the topological twist in a related context has been recently made in \cite{Pico:2026rji}.

%%%%%%%%%%%%%%
%%%%%%%%%%%%%%

\section{Global generalised structures and wrapped-M5 AdS$_5$ vacua} \label{sec:Subsectors}

%%%%%%%%%%%%%%
%%%%%%%%%%%%%%

In order to illustrate the discussion of section \ref{sec:4DSugra}, we will now particularise the $\cN=8$ truncation to specific subsectors that contain the M5-brane AdS$_5$ solutions of \cite{Maldacena:2000mw,Bah:2011vv,Bah:2012dg}. We will use the language of generalised $G$-structures, following \cite{Cassani:2019vcl,Cassani:2020cod,Josse:2025uro}.

%%%%%%%%%%%%%%

\subsection{Local and global generalised structures} \label{sec:SO3S}

%%%%%%%%%%%%%%

As shown in \cite{Cassani:2019vcl} building on \cite{Gauntlett:2007ma}, a consistent truncation of $D=11$ supergravity on a six-dimensional manifold $M_6$ to some $D=5$ gravitational theory always exists whenever $M_6$ is equipped with a constant-torsion generalised $G$-structure, with $G \subset \textrm{USp}(8)$. The resulting $D=5$ theory will be a (gauged, possibly matter coupled) $\cN$-extended supergravity when the $\bm{8}$ of USp$(8)$, in which the ExGG gravitini transform, yields $\cN$ singlets under the branching $\textrm{USp}(8) \supset G $. Equivalently, though in more geometric terms, this generalised $G$-structure is a reduction of the generic generalised USp(8)-structure associated to the generalised ExGG metric $G_{MN}(x,y)$. In particular, when $M_6$ is equipped with a GIS, {\it i.e.}~$G= \bm{1}$, the resulting truncation of $D=11$ supergravity on $M_6$ will be maximally supersymmetric. The $D=5$ $\cN=8$ consistent truncation of $D=11$ supergravity on the general BBBW manifolds $M_6 = \Sigma_2 \rtimes_{p,q} S^4$ of \cite{Bah:2011vv,Bah:2012dg} reported in section \ref{sec:TruncTwisted} above, and in \cite{Bhattacharya:2024tjw} for the specific $p=1$, $q=0$ MN2 \cite{Maldacena:2000mw} instance within this class, all fit in this scheme.

The MN-BBBW $\textrm{AdS}_5 \times (\Sigma_2 \rtimes_{p,q} S^4)$ wrapped M5-brane configurations \cite{Maldacena:2000mw,Bah:2011vv,Bah:2012dg} have been previously discussed in the context of E$_{6(6)}$ ExGG in \cite{Cassani:2019vcl,Cassani:2020cod}. The latter references identified a constant-torsion generalised U$(1)_{z}$-structure on $\Sigma_2 \rtimes_{p,q} S^4$, with $z$ defined in terms of $p$, $q$ through (\ref{eq:defz}) and, accordingly, various submaximal $D=11$ truncations to $D=5$ supergravity on those manifolds. More concretely, we will distinguish three relevant generalised U$(1)_{z}$-structures, depending on specific values of $z$.  For $|z|=1$, corresponding to a reduction of $D=11$ supergravity on MN2, the resulting supergravity is $D=5$ $\cN=4$ coupled to three vector multiplets \cite{Cassani:2019vcl}. This theory was also obtained from $D=11$ on MN2 by other methods in \cite{MatthewCheung:2019ehr}. For $z=0$, corresponding to a reduction of $D=11$ supergravity on MN1, the resulting supergravity is $D=5$ $\cN=2$ coupled to four vector multiplets and a hypermultiplet \cite{Cassani:2020cod} (see also \cite{Faedo:2019cvr} for a further subsector of this model). Finally, for generic $z$ away from those values, corresponding to a reduction of $D=11$ supergravity on generic hyperbolic BBBW, the resulting supergravity is $D=5$ $\cN=2$ coupled to two vector multiplets and a hypermultiplet \cite{Cassani:2020cod} (see also \cite{Szepietowski:2012tb} for a further subsector of this model and \cite{Gauntlett:2007sm} for a minimal subtruncation).

Since $\bm{1} \subset \textrm{U}(1)_{z}$, the GIS on $\Sigma_2 \rtimes_{p,q} S^4$ presented in section \ref{sec:TruncTwisted}, and in \cite{Bhattacharya:2024tjw} for $z=1$, arises as a further reduction of the generalised $\textrm{U}(1)_{z}$-structures of \cite{Cassani:2019vcl,Cassani:2020cod}. Consequently, the $D=5$ models of \cite{MatthewCheung:2019ehr,Cassani:2019vcl,Cassani:2020cod,Faedo:2019cvr,Szepietowski:2012tb} should all arise as subsectors of our maximally supersymmetric truncation. In \cite{Bhattacharya:2024tjw,Varela:2025xeb} that statement was indeed proved at the $D=5$ level: all those submaximal models do arise as consistent, U$(1)_{z}$-invariant subsectors of $D=5$ $\cN=8$ TCSO$(5,0,1;1)$-gauged supergravity. This is so even if $\textrm{U}(1)_{z}$ is not a subgroup of TCSO$(5,0,1;1)$\footnote{More generally, see \cite{Pico:2025cmc} for the conditions that render consistent an invariant subtruncation of maximal supergravity in the analogue $D=4$ case.}. In section \ref{sec:U1pqFrom11D} we will discuss how the generalised $\textrm{U}(1)_{z}$-structures of \cite{Cassani:2019vcl,Cassani:2020cod} can be recovered from our GIS (\ref{eq:TwistedFrame}), thereby recovering the consistent  truncations of those references at the $D=11$ level.

While our GIS is a further reduction of the generalised $\textrm{U}(1)_{z}$-structures of \cite{Cassani:2019vcl,Cassani:2020cod}, there is a crucial difference in character between them. The generalised $\textrm{U}(1)_{z}$-structures of \cite{Cassani:2019vcl,Cassani:2020cod} only depend on objects that extend globally on the six-manifold $\Sigma_2 \rtimes_{p,q} S^4$ and, for that reason, are globally defined in ExGG. On the contrary, our GIS is only locally defined on $\Sigma_2 \rtimes_{p,q} S^4$. This is because, as discussed in section~\ref{sec:TruncTwisted}, it is obtained by trading the unpunctured, hyperbolic Riemann surface $\Sigma_2$ with the group manifold $B_2$ of the non-abelian two-dimensional Lie algebra. While both manifolds, $B_2$ and $\Sigma_2$ are locally diffeomorphic, their equivalence does not extend globally. Indeed, $\Sigma_2$ can be compactified by a discrete group of isometries while $B_2$, being non-unimodular, cannot be compactified. The trombone gauging involved in the $D=5$ $\cN=8$ TCSO$(5,0,1;1)$ theory arises precisely because of the non-compactness of $B_2$. In section \ref{sec:GlobalSpecMN1} we will come back to this global versus local character and its implications for the KK spectrum.

%%%%%%%%%%%%%%

\subsection{The U$(1)_{z}$-invariant sector from $D=11$} \label{sec:U1pqFrom11D}

%%%%%%%%%%%%%%

Let us now make contact with the submaximal $D=11$ truncation on BBBW discussed in \cite{Cassani:2020cod}. For this purpose, our strategy will be simply to particularise the maximal truncation of section \ref{sec:TruncTwisted} to the U$(1)_{z}$-invariant sector of $D=5$ $\cN=8$ TCSO$(5,0,1;1)$ supergravity. This U$(1)_{z}$ is embedded in $\textrm{USp}(8) \subset \textrm{E}_{6(6)}$ via
{\setlength\arraycolsep{2pt}
\begin{eqnarray} \label{eq:branching}
\textrm{USp}(8) & \supset & \textrm{SU}(4) \times \textrm{U}(1)
\; \supset \;  \textrm{SO}(5) \times \textrm{U}(1)
\; \supset \;  \textrm{SU}(2)_+ \times \textrm{SU}(2)_- \times \textrm{U}(1) \nonumber \\[5pt]
& \; \supset \; & \textrm{U}(1)_+ \times \textrm{U}(1)_- \times \textrm{U}(1)
 \; \supset \; \textrm{U}(1)_+ \times \textrm{U}(1)_- \times \textrm{U}(1)_{z} \; , \nonumber \\[5pt]
&&  \textrm{with $\textrm{U}(1)_{z} \equiv z \,  \textrm{U}(1)_+ +  \, \textrm{U}(1)_- +\textrm{U}(1)$} \; ,
\end{eqnarray}
}and can be taken to be generated by
\begin{equation} \label{eq:U1Struc}
t_{p q} = -\tfrac{p}{p+q}  ( t_1{}^2 - t_2{}^1 ) +\tfrac{q}{p+q}  ( t_3{}^4 - t_4{}^3 ) - (\tilde{t}_1{}^2-\tilde{t}_2{}^1) \; ,
\end{equation}
with $p$, $q$ related to $z$ through (\ref{eq:defz}). The r.h.s.~of (\ref{eq:U1Struc})~again features the generators of E$_{6(6)}$ in the conventions of appendix A of \cite{Varela:2025xeb}, only with tildes over the $\textrm{SL}(2 , \mathbb{R}) \subset \textrm{E}_{6(6)}$ generators to distinguish them from the untilded $\textrm{SL}(6 , \mathbb{R}) \subset \textrm{E}_{6(6)}$ generators. By (\ref{eq:branching}), for generic $z$, $\textrm{U}(1)_{z}$ commutes with $\textrm{U}(1)_+ \times \textrm{U}(1)_-$ inside USp$(8)$. These factors are to be respectively identified with the flavour and R-symmetry of the $\cN=1$ BBBW SCFT. When $z=0$, the U$(1)_+$ factor is not involved in the diagonal, and U$(1)_0$ commutes with the enhanced $\textrm{SU}(2)_+ \times \textrm{U}(1)_-$ symmetry of MN1. For $z=1$ there are other enhancements corresponding to MN2. We will hereafter exclude the $z=1$ case, which has been covered in \cite{Bhattacharya:2024tjw}. 

The prescription of \cite{Cassani:2019vcl,Cassani:2020cod} entails retaining $D=5$ fields that result from expansion along the U$(1)_{z}$-invariants contained in various ExGG generalised bundles on the internal six-dimensional manifold $\Sigma_2 \rtimes_{p,q} S^4$ \cite{Bah:2011vv,Bah:2012dg} of interest. The gauge fields, in particular, descend from the generalised bundle in the $\overline{\bm{27}}_1$ of $\mathbb{R}^+ \times \textrm{E}_{6(6)}$. Under (\ref{eq:branching}), the $\overline{\bm{27}}_1$ gives three singlets for generic $z$. Taking advantage of our GIS $\hat{\tilde{E}}^M{}_{\uN} (y)$ on $\Sigma_2 \rtimes_{p,q} S^4$, (\ref{eq:TwistedFrame}) with (\ref{eq:E6GroupElement}), \eqref{eq:GenFrameSplit2}, \eqref{eq:GenFrameB2S4}, the corresponding generalised vectors $\hat{{\cal E}}_I{}^M(y) $, $I=0,1,2$, can be written as 
\begin{equation} \label{eq:GenVecBBBW}
\hat{{\cal E}}_I{}^M (y) = \hat{K}_I{}^{\uN} \, \hat{\tilde{E}}^M{}_{\uN}  (y) \; .
\end{equation}
Here, $\hat{K}_I{}^{\uN}$ are constant, U$(1)_{z}$-invariants defined by $\hat{K}_I{}^{\uM} \, (t_{p q})_{\uM}{}^{\uN} =0$, with $t_{p q}$ in (\ref{eq:U1Struc}). Splitting the index $\uM$ as for $\tilde{X}_{\uM}$ above (\ref{eq:XSymbolsTilde}), the non-vanishing components of these generalised vectors are
\begin{equation}
\hat{K}_0{}^{12} = 1 \; , \quad
\hat{K}_0{}^{34} = 1  \; , \quad 
\hat{K}_0{}^{56} =1 \; .
\end{equation}
Similarly, the retained $D=5$ scalars lie in the coset $\mathrm{C}_{\mathrm{E}_{6(6)}}(\textrm{U}(1)_z) / \mathrm{C}_{\mathrm{USp}(8)}(\textrm{U}(1)_z)$, where $\mathrm{C}_{K}(G)$ denotes the commutant of $G \subset K$ inside $K$ \cite{Cassani:2019vcl,Cassani:2020cod}. In the case at hand, some calculation reveals that 
\begin{equation} \label{eq:U1pqCoset}
\mathrm{C}_{\mathrm{E}_{6(6)}}(\textrm{U}(1)_z) / \mathrm{C}_{\mathrm{USp}(8)}(\textrm{U}(1)_z) \; = \; \textrm{SO}(1,1)^2 \times \frac{\textrm{SU}(2,1)}{\textrm{SU}(2) \times \textrm{U}(1)} \; .
\end{equation}
Altogether, the resulting field content for generic $z$ is compatible with $D=5$ $\cN=2$ supergravity coupled to two vector multiplets and a hypermultiplet, with scalar manifold (\ref{eq:U1pqCoset}), in agreement with \cite{Cassani:2020cod}. The generalised metric $G_{MN}$ is also U$(1)_{z}$-invariant and is determined exclusively by the generalised vectors $\hat{{\cal E}}_I{}^M$ in (\ref{eq:GenVecBBBW}) \cite{Cassani:2020cod}. The embedding of the $D=5$ scalars into the internal metric, $g_{mn}$, and warp factor, $e^{2\tilde{\Delta}}$, in the standard formulation of \cite{Cremmer:1978km} can then be extracted from the expressions
\begin{equation} \label{eq:ExtractComps}
(G^{-1})^{mn} = e^{2\tilde{\Delta}} \, g^{mn} \;  , \qquad 
e^{9\tilde{\Delta}} = \big( \textrm{det} (G^{-1})^{MN} \big)^{\frac{1}{18}} \big(\textrm{det} (G^{-1})^{mn} \big)^{\frac{1}{9}} \; . 
\end{equation}

Alternatively, we can make use of our full $\cN=8$ machinery to uplift the U$(1)_{z}$-invariant sector of $D=5$ $\cN=8$ TCSO$(5,0,1;1)$-gauged supergravity. By truncating the five-dimensional maximal supergravity to its U$(1)_{z}$-invariant sector, we can uplift the latter using the simpler generalised Scherk--Schwarz formulae (\ref{eq:GenSS}), rather than the more complicated U$(1)_{z}$-invariant formulae of \cite{Cassani:2020cod}.  Unlike its parent $D=5$ $\cN=8$ theory, this subsector is described by a Lagrangian as the $\textrm{U}(1)_{z}$-invariant trombone components of the embedding tensor (\ref{eq:XSymbolsTilde}) vanish \cite{Varela:2025xeb}. The gravity-scalar part of this Lagrangian reads \cite{Cassani:2020cod}, in the conventions of \cite{Varela:2025xeb}, 
{\setlength\arraycolsep{2pt}
\begin{eqnarray}	\label{eq:Lagrangian}
{\cal L} &=&  R \, \textrm{vol}_5 - 12 ( d\varphi_0 )^2   - 4 ( d\varphi_1 )^2  -  2(D\phi_0 )^2 - \tfrac{1}{2} \, e^{4 \phi_0} \,   \big( Da +  \tfrac{1}{2}  ( \xi D \tilde{\xi} - \tilde{\xi} D \xi  ) \big)^2 \nonumber   \\[5pt]
&&  -  \tfrac{1}{2} \, e^{2 \phi_0} \,  ( D\xi )^2    -  \tfrac{1}{2} \, e^{2 \phi_0} \,   (D\tilde{\xi} )^2 - V \, \textrm{vol}_5  + \ldots   \ ,
\end{eqnarray}
}where $ ( d\varphi )^ 2 \equiv d\varphi  \wedge * d\varphi $, etc., the ellipses denote contributions from other supergravity fields, which we will ignore subsequently, and the scalar potential is
{\setlength\arraycolsep{2pt}
\begin{eqnarray} \label{eq:BBBWPot}
g^{-2} \, V &= &  \tfrac12 \, e^{4\phi_0 + 8 \varphi_0} + 2e^{2\phi_0 -4 \varphi_0}  - 2 e^{2( \phi_0 + \varphi_0-\varphi_1) }- 2 e^{2( \phi_0 + \varphi_0+ \varphi_1) } -4 e^{ -4 \varphi_0}  \nonumber \\[5pt]
&& +\tfrac12 \, e^{2\phi_0-4\varphi_0} ( \xi^2 + \tilde{\xi}^2 ) \big( e^{-4 \varphi_1} + e^{4 \varphi_1} +2 e^{2\phi_0} -2 \big) 
 \nonumber \\[5pt]
&& +\tfrac18 \, e^{4 (\phi_0- \varphi_0 + \varphi_1 ) } ( 1+ z - \xi^2 - \tilde{\xi}^2 )^2
+\tfrac18 \, e^{4 (\phi_0- \varphi_0 - \varphi_1 ) } ( 1 - z - \xi^2 - \tilde{\xi}^2 )^2 \; .
\end{eqnarray}
}Here, we have set $g_1 = g_2 \equiv g$ (and $R=R_\Sigma$ below by (\ref{eq:CouplingsRadii})) for simplicity and have used $z$ defined in (\ref{eq:defz}). In (\ref{eq:Lagrangian}), (\ref{eq:BBBWPot}), $\varphi_0$, $\varphi_1$, $\phi_0$, $a$, $\xi$, $\tilde{\xi}$ denote the $\textrm{U}(1)_{z}$-invariant scalars contained in the $D=5$ $\cN=8$ supergravity: the former two parametrise the $\textrm{SO}(1,1)^2$ factor of (\ref{eq:U1pqCoset}), and the remaining four the rightmost factor. Regarding (\ref{eq:U1pqCoset}) as a submanifold of $\textrm{E}_{6(6)}/ \textrm{USp}(8)$, a coset representative can be taken, in Iwasawa gauge, as
\begin{equation} \label{eq:SugraSector}
\cV = e^{\varphi_0 H_0} \, e^{-\sqrt{2} \, \varphi_1 B}  \, e^{-\frac{1}{\sqrt{2}} ( a E_2 - \xi E_{11} +\tilde{\xi} E_{12} ) } \, e^{-\phi_0 H_1} \; .
\end{equation}
The generators $H_0$, $B$, $E_2$, $E_{11}$ and $E_{12}$ are in $\mathrm{C}_{\mathrm{E}_{6(6)}}(\textrm{U}(1)_{z})$, and have been explicitly given in (4.6) of \cite{Varela:2025xeb}. 

Now, the $D=11$ uplift of the U$(1)_{z}$-invariant metric-scalar sector can be obtained by bringing the scalar matrix $M \equiv \cV \cV^{\textrm{T}}$, with $\cV$ in (\ref{eq:SugraSector}), and the generalised frame $\hat{\tilde{E}}^M{}_{\uN} (y)$ in (\ref{eq:TwistedFrame}) with (\ref{eq:E6GroupElement}), \eqref{eq:GenFrameSplit2}, \eqref{eq:GenFrameB2S4} to the generalised 
Scherk--Schwarz expression (\ref{eq:GenSS}) for the inverse generalised metric, $(G^{-1})^{MN}$. Finally, the ordinary metric and warp factor in the standard $D=11$ supergravity formulation of \cite{Cremmer:1978km} can then be obtained with the help of (\ref{eq:ExtractComps}). After considerable massaging, the $D=11$ metric finally takes on the form
{\setlength\arraycolsep{2pt}
\begin{eqnarray} \label{eq:11DMetric}
ds_{11}^2 &=& e^{\frac23 (\phi_0-\varphi_0)}  \Delta_0^{1/3} \, \Big\{ ds_5^2 + R^2 \, e^{-2\phi_0+4\varphi_0} \frac{d x^2+d y^2 }{y^2}  \\[5pt]
&& +\tfrac14 R^2\, \Delta_0^{-1} \, \Big[ 
4 \, e^{-4\varphi_0}
\tilde{\Delta}_0 \,d\zeta^2 + \sin^2\zeta \,   \big[ \Delta_1 \big( D\theta^2+ \sin^2\theta\, D\phi^2 \big) + \Delta_2  \big( D\psi+\cos\theta\,D\phi \big)^2 \big] \nonumber 
\\[5pt]
&& + \left(e^{-2\varphi_1}- e^{2\varphi_1}\right) \sin \zeta \sin\theta \big[
2  \cos \zeta \,d\zeta\,D\theta - \sin\zeta \, \sin \theta D\phi\,\big( D\psi+\cos\theta\,D\phi \big) \big]  \Big] \Big\}  \; .\nonumber 
\end{eqnarray}
}The standard metric (\ref{eq:B2Geom}) on $\Sigma_2$ naturally appears here, while the last two lines on the r.h.s.~correspond to a metric on $S^4$, parametrised by angles $(\zeta , \theta, \phi , \psi)$ defined in appendix~\ref{sec:S4Conv}. In (\ref{eq:11DMetric}), we have defined the following functions of the $D=5$ scalars and the $S^4$ angles
\begin{eqnarray} \label{eq:MetricFuncs}
& \Delta_1 =e^{-2\varphi_1}\cos^2\tfrac{\theta}{2}
  +e^{2\varphi_1}\sin^2\tfrac{\theta}{2} \; , \quad
\Delta_2 =e^{-2\varphi_1}\sin^2\tfrac{\theta}{2}
  +e^{2\varphi_1}\cos^2\tfrac{\theta}{2} \; , \quad  \nonumber \\[5pt]
& \Delta_0 =e^{-4\varphi_0}\Delta_1 \sin^2\zeta\,
 +e^{2\phi_0+2\varphi_0}\cos^2\zeta \; , \quad
\tilde{\Delta}_0 =
 e^{-2\phi_0-2\varphi_0} \sin^2\zeta  + e^{4\varphi_0}\Delta_2\cos^2\zeta  . \quad 
\end{eqnarray}
Finally, the scalar-dependent covariant derivatives
{\setlength\arraycolsep{2pt}
\begin{eqnarray} \label{eq:AngCovDer}
D\theta
&=& d\theta -  \left(\xi\cos\psi-\tilde{\xi}\sin\psi\right) \frac{d x}{y} -  \left(\tilde{\xi} \cos\psi+\xi\sin\psi\right) \frac{d y}{y},
\nonumber \\[5pt]
D\phi &=& d\phi - \left[ z+  \frac{\xi\sin\psi+\tilde{\xi} \cos\psi}{\sin\theta} \right] \frac{d x}{y} + \frac{\xi\cos\psi-\tilde{\xi}\sin\psi}{\sin\theta} \frac{d y}{y},
\\[5pt]
D\psi &=& d\psi - \left[ 1-  \cot\theta \left(\xi\sin\psi + \tilde{\xi}\cos\psi\right) \right] \frac{d x}{y} + \cot\theta \left(\tilde{\xi}\sin\psi-\xi\cos\psi\right) \frac{d y}{y}. \nonumber 
\end{eqnarray}
}implement the fibration of $S^4$ over $\Sigma_2$.

%%%%%%%%%%%%%%

\subsection{Recovering BBBW} \label{sec:BBBW}

%%%%%%%%%%%%%%

At fixed $z$, the scalar potential (\ref{eq:BBBWPot}) attains an AdS vacuum located at \cite{Varela:2025xeb}
{\setlength\arraycolsep{0pt}
\begin{eqnarray} \label{eq:BBBWvac}
& \textrm{BBBW: } \qquad  e^{12\varphi_0} = \frac{ z^2 \big( 2 + \sqrt{1+3z^2} \big)^2}{ \big( 3z-1 + \sqrt{1 + 3z^2 } \big) \big( 3z + 1 - \sqrt{1 + 3z^2}  \big) } \; , \qquad 
e^{4\varphi_1} = \frac{  3z-1 + \sqrt{1 + 3z^2 } }{ 3z + 1 - \sqrt{1 + 3z^2 } } \; ,  \nonumber \\[5pt]
& \qquad \qquad e^{-2\phi_0} = \tfrac12 +\tfrac14  \sqrt{1 + 3z^2 } \; , \quad  \xi = \tilde{\xi} =0 \; , \quad
L^2 = \left( \frac{ 9 z^2 -1 + (1+3z^2)^{3/2}}{4z^2} \right)^{2/3} g^{-2} . \qquad 
\end{eqnarray}
}For $z=0$ ($p=q$) and $z=1$ ($q=0$), (\ref{eq:BBBWvac}) becomes
\begin{eqnarray}
\label{eq:MN1vac}
\textrm{MN1} & : & \quad z=0 \; , \quad e^{2 \phi_0} = \tfrac43 \; , \quad 
\varphi_0 = \varphi_1 = \xi = \tilde{\xi} =0 \; , \quad L^2 = \tfrac94 \, g^{-2} \; , \\[5pt]
\label{eq:MN2vac}
\textrm{MN2} & : &  \quad z=1 \; , \quad e^{12 \varphi_0 } = e^{4 \varphi_1 }  = 2 \; , \quad 
\phi_0 = \xi = \tilde{\xi} =0 \; , \quad L^2 = 2^{4/3}  \, g^{-2} \; .
\end{eqnarray}
The AdS radius, $L^2 = -12/V$ with $V$ in (\ref{eq:BBBWPot}) evaluated on the scalar vacuum values, has also been given in these equations. By the consistency of the U$(1)_{z}$-invariant subtruncation, these are also vacua of the full $D=5$ $\cN=8$ TCSO$(5,0,1;1)$ supergravity. Within the latter, the generic vacuum (\ref{eq:BBBWvac}) spontaneously breaks $\cN=8$ supersymmetry down to $\cN=2$, and the TCSO$(5,0,1;1)$ gauge symmetry to the subgroup $\textrm{U}(1)_+ \times \textrm{U}(1)_-$ defined in (\ref{eq:branching}). The $\textrm{U}(1)_-$ factor is the R-symmetry and $\textrm{U}(1)_+$ is flavour. The specific vacuum (\ref{eq:MN1vac}) also preserves $\cN=2$ supersymmetry, but the bosonic symmetry is enhanced to the $\textrm{SU}(2)_+ \times \textrm{U}(1)_-$ also defined in (\ref{eq:branching}). In this case, $\textrm{U}(1)_-$ still is the R-symmetry and the flavour symmetry is enhanced. Finally, the vacuum (\ref{eq:MN2vac}) has supersymmetry augmented to $\cN=4$, no flavour, and R-symmetry $\textrm{SU}(2) \times \textrm{U}(1)_-$, with $\textrm{SU}(2)$ the diagonal subgroup of $\textrm{SU}(2)_+ \times \textrm{SU}(2)_-$ defined in the first line of (\ref{eq:branching}). Our notation is such that $\cN=2$ and $\cN=4$ AdS$_5$ vacua are respectively dual to $\cN=1$ and $\cN=2$ SCFTs.

The $z=1$ vacuum (\ref{eq:MN2vac}) was shown in \cite{Bhattacharya:2024tjw} to uplift to the MN2 $D=11$ solution \cite{Maldacena:2000mw}. It was argued in \cite{Varela:2025xeb} that the other vacua, (\ref{eq:BBBWvac}), (\ref{eq:MN1vac}), uplift to the indicated $D=11$ $\textrm{AdS}_5 \times (\Sigma_2 \rtimes_{p,q} S^4)$ solutions: BBBW \cite{Bah:2011vv,Bah:2012dg} and MN1 \cite{Maldacena:2000mw}, all with an unpunctured, hyperbolic Riemann surface $\Sigma_2$. Using the uplift formulae of section \ref{sec:U1pqFrom11D}, we can now show that this is the case: at the $D=5$ vacuum (\ref{eq:BBBWvac}), the uplifted metric (\ref{eq:11DMetric}) indeed reduces to that of the hyperbolic BBBW solution \cite{Bah:2011vv,Bah:2012dg}
{\setlength\arraycolsep{2pt}
\begin{eqnarray} \label{eq:BBBWMetric}
ds_{11}^2 &=& X_0^{-2} \,  \bar{\Delta}^{1/3} \, ds^2 (\textrm{AdS}_5)  +  \bar{\Delta}^{1/3} e^{2g_0} \frac{d x^2+d y^2 }{y^2}  \\[5pt]
&& +\tfrac14 \, \bar{\Delta}^{-2/3} \, \Big[ 
X_0^{-1} d\mu_0^2 +  X_1^{-1} \big( d\mu_1^2+  \mu_1^2(d\chi_1+A_1)^2 \big)+  X_2^{-1} \big( d\mu_2^2+  \mu_2^2(d\chi_2+A_2)^2 \big) \Big] .\nonumber
\end{eqnarray}
}Here, $\chi_1$, $\chi_2$, $\mu_0$, $\mu_1$, $\mu_2$, with $\mu_0^2+\mu_1^2+\mu_2^2 =1$ parametrise the $S^4$; the one-forms
\begin{eqnarray} \label{eq:BBBWFib}
A_1 = \tfrac{z+1}{2} \tfrac{dx}{y} \; , \qquad 
A_2 = \tfrac{z-1}{2} \tfrac{dx}{y}
\end{eqnarray}
fibre $S^4$ over the Riemann surface; $X_0$, $X_1$, $X_2$, $e^{2g_0}$ are the $z$-dependent constants
{\setlength\arraycolsep{0pt}
\begin{eqnarray} \label{eq:BBBWConst}
&  X_1^5 =  \tfrac{  1+7z+7z^2+33z^3  -(1+4z+19z^2)\sqrt{1+3z^2}  }{  4z(1 -z)^2 } \; , \quad 
  X_1X_2^{-1} =  \tfrac{1+z}{2z+\sqrt{1+3z^2}}  \; , \quad 
  X_0=(X_1X_2)^{-2} \; ,  \nonumber \\[5pt]
&  e^{2g_0}=\frac18X_1X_2\left[(1-z)X_1+(1+z)X_2\right] \; ;
\end{eqnarray}
}and, finally, the warp factor takes on the form
\begin{equation}
\bar{\Delta} = X_0 \mu_0^2 +X_1 \mu_1^2 +X_2 \mu_2^2 \; ,
\end{equation}
in terms of the constrained coordinates $\mu_0$, $\mu_1$, $\mu_2$ on the $S^4$ and the constants (\ref{eq:BBBWConst}). 

In order to see that (\ref{eq:11DMetric}) with (\ref{eq:BBBWvac}) does reproduce (\ref{eq:BBBWMetric}), it is helpful to note the relations
\begin{equation}
X_0 = e^{\frac{8}{5}(\phi_0+3\varphi_0)} \; , \;\;
X_1 = e^{-\frac{2}{5}(\phi_0+3\varphi_0+5\varphi_1)} \; , \;\;
X_2 = e^{-\frac{2}{5}(\phi_0+3\varphi_0-5\varphi_1)}\; , \;\;
e^{2g_0} = \tfrac14 \, e^{-\frac{6}{5}(\phi_0+2\varphi_0)} \; , 
\end{equation}
between the quantities (\ref{eq:BBBWConst}) and the $D=5$ vacuum scalars (\ref{eq:BBBWvac}). Also, the $S^4$ angles that appear in (\ref{eq:11DMetric}) and (\ref{eq:BBBWMetric}) need to be related through  (\ref{eq:AltS4Angles}). In particular, evaluated on (\ref{eq:BBBWvac}), the warp factors $\Delta_0$ and $\bar{\Delta}$ become related as
\begin{equation}
\bar{\Delta} = e^{-\frac{2}{5}(\phi_0-7\varphi_0)} \Delta_0 \; ,
\end{equation}
and the fibrations that appear in (\ref{eq:BBBWMetric}), (\ref{eq:BBBWFib}) become
\begin{equation}
d\chi_1 +A_1 = -\tfrac12 (D\phi + D\psi) \; , \qquad
d\chi_2 +A_2 = \tfrac12 (D\phi - D\psi) \; ,
\end{equation}
in terms of the covariant derivatives (\ref{eq:AngCovDer}) with $\xi=\tilde{\xi}=0$.

%%%%%%%%%%%%%%

\subsection{Recovering MN1} \label{sec:MN1}

%%%%%%%%%%%%%%

The $D=5$ vacuum (\ref{eq:MN1vac}) similarly uplifts to the MN1 solution. Indeed, evaluated on (\ref{eq:MN1vac}), the functions (\ref{eq:MetricFuncs}) reduce to $\Delta_1= \Delta_2 =1$, $3  \Delta_0 = 4  \tilde{\Delta}_0 = 3+\cos^2 \zeta$, and the 
$D=11$ metric (\ref{eq:11DMetric}) collapses to 
{\setlength\arraycolsep{2pt}
\begin{eqnarray} \label{eq:11DMetricMN1}
ds_{11}^2 &=& 2^{\frac23} \cdot 3^{-\frac23}  ( 3+\cos^2 \zeta )^{\frac13} \, \Big\{ L^2 ds^2 (\textrm{AdS}_5)  \nonumber   \\[3pt]
&& 
+\tfrac34 R^2\, \Big[ \frac{d x^2+d y^2 }{y^2} + 
d\zeta^2 + \frac{\sin^2\zeta}{3+\cos^2 \zeta}  \, \big( \sigma_1^2 + \sigma_2^2 + (\sigma_3 + \upsilon )^2 \big) \Big] \Big\} \; ,
\end{eqnarray}
}with $\sigma_1$, $\sigma_2$, $\sigma_3$ the left-invariant forms (\ref{eq:SU2LIForms}) on the $\textrm{SU}(2)_+ $ defined in (\ref{eq:branching}), and $\upsilon$ the spin connection (\ref{eq:SpinCon}) on $\Sigma_2$. This is the MN1 metric \cite{Maldacena:2000mw}, as presented in \cite{Cassani:2020cod}. This metric exhibits an explicit $\textrm{SU}(2)_+ \times \textrm{U}(1)_-$ isometry, in agreement with the symmetry of the vacuum (\ref{eq:MN1vac}) within $D=5$ $\cN=8$ TCSO$(5,0,1;1)$ supergravity.

%%%%%%%%%%%%%%%
%%%%%%%%%%%%%%%

\section{The U$(1)_0$-invariant universal spectrum of MN1} \label{sec:Spectra}

%%%%%%%%%%%%%%%
%%%%%%%%%%%%%%%

We now specialise the maximal truncation to the MN1 vacuum and determine the universal U$(1)_0$-invariant sector of its light operator spectrum at arbitrary KK level.

%%%%%%%%%%%%%%%
\subsection{From the local putative spectrum to the global MN1 sector} \label{sec:PutUniv}

The maximally supersymmetric consistent truncation of $D=11$ supergravity on the BBBW twisted geometries $\Sigma_2 \rtimes_{p,q} S^4$, introduced in section~\ref{sec:TruncTwisted}, allows the MN1 background to be analysed using the ExFT spectral machinery of \cite{Malek:2019eaz,Malek:2020yue,Varela:2020wty,Cesaro:2020soq}, extended to trombone gaugings in \cite{Pico:2026kch}. For the truncation at hand, however, the GIS constructed in section~\ref{sec:TruncTwisted} exists only locally on $\Sigma_2 \rtimes_{p,q} S^4$, as discussed in section~\ref{sec:SO3S}. Accordingly, the spectrum produced directly by this construction should likewise be regarded, a priori, as only locally defined. By contrast, the generalised U$(1)_0$-structure on the MN1 bundle, originally constructed in \cite{Cassani:2020cod} and reviewed in section~\ref{sec:Subsectors}, is globally well defined. We shall therefore interpret the local spectrum generated by the ExFT methods as a reservoir of candidate physical modes and refer to it as the putative spectrum. Requiring U$(1)_0$ invariance selects a guaranteed globally defined, $\Sigma_2$-constant subsector. Both the putative spectrum and this global subsector are universal in the sense that they are insensitive to the particular compact quotient $\Sigma_2=H^2/\Gamma$.

At KK level $k$, the individual states in the putative spectrum transform in representations $\bm{r} \times [k0]$ of $\textrm{USp}(8) \times \textrm{SO}(5)$. The representation $\bm{r}$ is inherited from the $k=0$ state at the base of the corresponding tower, viewed as a linearised field of $D=5$ $\cN=8$ supergravity. Thus, $\bm{r}$ is respectively $\bm{1}$, $\overline{\bm{8}}$, $\bm{\overline{27}}$, $\bm{27}$, $\overline{\bm{48}}$ or $\bm{42}$ for gravitons, gravitini, vectors, two-forms, spin-$1/2$ fermions or scalars, as reviewed in section~\ref{sec:FieldContentET}. Also, $[k0]$ is the rank-$k$ symmetric traceless representation of SO$(5)$. The KK mass matrices appropriate to trombone gaugings were obtained in \cite{Pico:2026kch} and are collected for reference in appendix~\ref{sec:KKMassMat}. They depend on the scalar matrix $M_{MN}=(\cV\cV^{\textrm{T}})_{MN}$ associated with the coset representative (\ref{eq:SugraSector}), evaluated at the MN1 vacuum (\ref{eq:MN1vac}). They also involve the TCSO$(5,0,1;1)$ embedding tensor (\ref{eq:XSymbolsTilde}) with (\ref{eq:Charges1}), (\ref{eq:ExtraCompsET}), together with the SO$(5)$ generators acting in the representation $\oplus_{k=0}^{\infty}[k0]$,
\begin{equation} \label{eq:TransCurlyT}
(\tilde{\cT}_{\uM})^\Sigma{}_{\Lambda}=U_{\uM}{}^{\uN}(\cT_{\uN})^\Sigma{}_{\Lambda}\; ,
\end{equation}
written in the duality frame determined by (\ref{eq:XSymbolsTilde}). In (\ref{eq:TransCurlyT}), $U_{\uM}{}^{\uN}$ denotes the transformation (\ref{eq:E7GroupElement}), while
\begin{equation} \label{eq:TMS4}
(\cT_{\uM})^\Sigma{}_{\Lambda}=\{0,\,(\cT_{ij})^h{}_{\ell},\,2(\cT_{ij})^{\{h_1}{}_{\{\ell_1}\delta^{h_2\}}_{\ell_2\}},\,\ldots,\,k(\cT_{ij})^{\{h_1}{}_{\{\ell_1}\delta^{h_2}_{\ell_2}\cdots\delta^{h_k\}}_{\ell_k\}},\,\ldots\}\; .
\end{equation}
Here, only the $(\bm{1},\bm{10})$ components of the fundamental E$_{6(6)}$ index $\uM$ are non-vanishing, and $(\cT_{ij})^h{}_{\ell}\equiv2\delta^h{}_{[i}\delta_{j]\ell}$ are the SO$(5)$ generators in the fundamental representation. One may verify that the generators (\ref{eq:TransCurlyT}) obey
\begin{equation} \label{eq:LieAlgHarm}
[\tilde{\cT}_{\uM},\tilde{\cT}_{\uN}]=-R\,\tilde{X}_{[\uM\uN]}{}^{\uP}\tilde{\cT}_{\uP}\; ,\qquad
\hat{\tilde{E}}^M{}_{\uN}(y)\,\partial_M\cY_{\Lambda}(y)=R^{-1}(\tilde{\cT}_{\uN})^\Sigma{}_{\Lambda}\cY_{\Sigma}(y)\; .
\end{equation}
Here, the scalar harmonic basis $\cY^\Lambda$ can be chosen, as in appendix~\ref{sec:GravOp}, as 
\begin{equation} \label{eq:harmS4}
\cY^\Lambda=\{1,\,y^i,\,y^{\{i_1}y^{i_2\}},\,\ldots,\,y^{\{i_1}\cdots y^{i_k\}},\,\ldots\}\; ,
\end{equation}
namely the $S^4$ spherical harmonics. The coordinates $y^i$ obey $\delta_{ij}y^iy^j=1$, and curly brackets denote traceless symmetrisation so that $y^{\{i_1}\cdots y^{i_k\}}$ lies in the SO$(5)$ representation $[k0]$.

The resulting putative KK spectrum is discrete and fully diagonalisable over the real numbers, even though the mass matrices are non-symmetric and the underlying maximal GIS is only locally defined. As already emphasised, locality arises because the compact surface $\Sigma_2$ is replaced by the non-unimodular group manifold $B_2$, which induces the trombone gauging. In the construction above the harmonic coefficients are taken to be constant along $\Sigma_2$, or equivalently along $B_2$. Such coefficients extend globally on the MN1 bundle precisely when the complete fluctuation is a U$(1)_0$ singlet. The individual putative eigenstates assemble into representations of $\textrm{SU}(2,2|1)\times\textrm{SU}(2)_+$, see appendix~\ref{sec:PutSpec}.

Operationally, the global sector is obtained by extracting the U$(1)_0$ singlets in the branching of the full product $\bm{r}\times[k0]$ supporting each putative tower. The relevant chain begins with
{\setlength\arraycolsep{0pt}
\begin{eqnarray} \label{eq:branchingKK}
\textrm{USp}(8)\times\textrm{SO}(5)&\supset&\textrm{SU}(4)\times\textrm{U}(1)\times\textrm{SO}(5)
\supset\textrm{SO}(5)\times\textrm{U}(1)\times\textrm{SO}(5)
\supset\textrm{SO}(5)\times\textrm{U}(1) \; ,  \qquad
\end{eqnarray}
}followed by $\textrm{SO}(5)\times\textrm{U}(1)\supset\textrm{SU}(2)_+\times\textrm{SU}(2)_-\times\textrm{U}(1)$ and the final U$(1)_0$ embedding in (\ref{eq:branching}). The final SO$(5)$ factor in (\ref{eq:branchingKK}) is the diagonal subgroup of the two SO$(5)$ factors at the preceding stage. The surviving global modes organise themselves into representations of $\textrm{SU}(2,2|1)\times\textrm{SU}(2)_+$, as we will discuss in section~\ref{sec:GlobalSpecMN1}. Appendix~\ref{sec:GravOp} gives an independent eleven-dimensional derivation for the graviton tower and explains how further global modes with non-trivial $\Sigma_2$ dependence are governed by weighted Maass operators.

\subsection{Global MN1 spectrum} \label{sec:GlobalSpecMN1}

Our principal spectral result is that the complete U$(1)_0$-invariant, $\Sigma_2$-constant universal MN1 spectrum can be given in closed form at arbitrary KK level. All superconformal-primary dimensions are specialisations of a single expression, and the full sector consists of the graviton, gravitino and vector-multiplet towers, in representations of the SU$(2)_+$ flavour symmetry, displayed below.

We find that the dimension of the superconformal primary of an $\textrm{SU}(2,2|1)$ multiplet at KK level $k$, with Lorentz spins $j_1$, $j_2$, SU$(2)_+$ flavour spin $\ell$ and U$(1)_-$ charge $n$, is:
\begin{equation} \label{eq:MN1DimFormula}
E_{k \ell n j_1j_2} = 1 + \sqrt{7- 2j_1(j_1+1) - 2j_2(j_2+1) +3 k(k+3) + \tfrac34 n^2 -3 \ell (\ell+1) }\,.
\end{equation}
From (\ref{eq:MN1DimFormula}) we also find it helpful to define the graviton, gravitino, neutral-vector and charged-vector dimensions:
{\setlength\arraycolsep{2pt}
\begin{eqnarray} \label{eq:MN1DimFormulaPart}
E^{(\textrm{grav})}_{k \ell} &=& E_{k \ell 0 \frac12\frac12} = 1 + \sqrt{4+3 k(k+3) -3 \ell (\ell+1) } \; , \nonumber \\[4pt]
E^{(\textrm{gino})}_{k \ell} &=& E_{k \ell 1 \frac12 0} = 1 + \sqrt{\tfrac{25}{4} +3 k(k+3) -3 \ell (\ell+1) } \; , \\[4pt]
E^{(\textrm{vec},0)}_{k \ell}  &=& E_{k \ell 0 0 0} = 1 + \sqrt{7 +3 k(k+3) -3 \ell (\ell+1) } \; , \nonumber \\[4pt]
E^{(\textrm{vec},2)}_{k \ell} &=& E_{k \ell 2 0 0} = 1 + \sqrt{10 +3 k(k+3) -3 \ell (\ell+1) } \; . \nonumber
\end{eqnarray}
}The superscripts $0, 2$ on $E^{(\textrm{vec},0)}_{k\ell}$ and $E^{(\textrm{vec},2)}_{k\ell}$ denote $n=0$ and $|n|=2$. Introduce also
\begin{equation} \label{eq:QNDefs}
m \equiv \left\lfloor \frac{k}{2} \right\rfloor \; ,  \qquad
h \equiv \left\lceil\frac{k}{2} \right\rceil \; ,
\end{equation}
in terms of the usual floor and ceiling functions. With these definitions, the U$(1)_0$-invariant spectrum of $\textrm{SU}(2,2|1)\times\textrm{SU}(2)_+$ multiplets of the MN1 solution is as follows.

\begin{table}[p]
\centering
\begingroup
\scriptsize
\setlength{\tabcolsep}{4pt}
\renewcommand{\arraystretch}{1.05}

% Fit the width, then reduce the height only if necessary.
\setbox0=\hbox{%
\resizebox{\textwidth}{!}{%
\begin{tabular}{l | l}

\hline
\hline
\textbf{Multiplet} & $k=4$ \\[2pt]
\hline

Graviton
&
$L\bar{L}
 \left[1+2\sqrt{22};\tfrac12,\tfrac12;0\right]
 \otimes[0]
 \;\oplus\;
 L\bar{L}
 \left[1+\sqrt{82};\tfrac12,\tfrac12;0\right]
 \otimes[1]$
\\[1pt]
&
$\oplus\;
 L\bar{L}
 \left[1+\sqrt{70};\tfrac12,\tfrac12;0\right]
 \otimes[2]$
\\[2pt]

\hline

Gravitino
&
$\Big(
 L\bar{L}
 \left[1+\tfrac{\sqrt{337}}{2};0,\tfrac12;1\right]
 \oplus
 L\bar{L}
 \left[1+\tfrac{\sqrt{337}}{2};\tfrac12,0;-1\right]
 \Big)\otimes[1]$
\\[1pt]
&
$\oplus\;
 \Big(
 L\bar{L}
 \left[1+\tfrac{\sqrt{337}}{2};0,\tfrac12;-1\right]
 \oplus
 L\bar{L}
 \left[1+\tfrac{\sqrt{337}}{2};\tfrac12,0;1\right]
 \Big)\otimes[1]$
\\[1pt]
&
$\oplus\;
 \Big(
 L\bar{L}
 \left[\tfrac{19}{2};0,\tfrac12;1\right]
 \oplus
 L\bar{L}
 \left[\tfrac{19}{2};\tfrac12,0;-1\right]
 \Big)\otimes[2]$
\\[2pt]

\hline

Vector
&
$2L\bar{L}
 \left[1+\sqrt{91};0,0;0\right]
 \otimes[0]
 \;\oplus\;
 2L\bar{L}
 \left[1+\sqrt{85};0,0;0\right]
 \otimes[1]$
\\[1pt]
&
$\oplus\;
 L\bar{L}
 \left[1+\sqrt{73};0,0;0\right]
 \otimes[2]
 \;\oplus\;
 L\bar{L}
 \left[1+\sqrt{55};0,0;0\right]
 \otimes[3]$
\\[1pt]
&
$\oplus\;
 \Big(
 L\bar{L}
 \left[1+\sqrt{76};0,0;2\right]
 \oplus
 L\bar{L}
 \left[1+\sqrt{76};0,0;-2\right]
 \Big)\otimes[2]$
\\[2pt]

\hline
\hline
& $k=3$ \\[2pt]
\hline

Graviton
&
$L\bar{L}
 \left[1+\sqrt{58};\tfrac12,\tfrac12;0\right]
 \otimes[0]
 \;\oplus\;
 L\bar{L}
 \left[1+2\sqrt{13};\tfrac12,\tfrac12;0\right]
 \otimes[1]$
\\[2pt]

\hline

Gravitino
&
$\Big(
 L\bar{L}
 \left[1+\tfrac{\sqrt{217}}{2};0,\tfrac12;1\right]
 \oplus
 L\bar{L}
 \left[1+\tfrac{\sqrt{217}}{2};\tfrac12,0;-1\right]
 \Big)\otimes[1]$
\\[1pt]
&
$\oplus\;
 \Big(
 L\bar{L}
 \left[1+\tfrac{\sqrt{217}}{2};0,\tfrac12;-1\right]
 \oplus
 L\bar{L}
 \left[1+\tfrac{\sqrt{217}}{2};\tfrac12,0;1\right]
 \Big)\otimes[1]$
\\[1pt]
&
$\oplus\;
 \Big(
 L\bar{L}
 \left[\tfrac{15}{2};0,\tfrac12;1\right]
 \oplus
 L\bar{L}
 \left[\tfrac{15}{2};\tfrac12,0;-1\right]
 \Big)\otimes[2]$
\\[2pt]

\hline

Vector
&
$2L\bar{L}
 \left[1+\sqrt{61};0,0;0\right]
 \otimes[0]
 \;\oplus\;
 L\bar{L}
 \left[1+\sqrt{55};0,0;0\right]
 \otimes[1]$
\\[1pt]
&
$\oplus\;
 L\bar{L}
 \left[1+\sqrt{43};0,0;0\right]
 \otimes[2]$
\\[2pt]

\hline
\hline
& $k=2$ \\[2pt]
\hline

Graviton
&
$L\bar{L}
 \left[1+\sqrt{34};\tfrac12,\tfrac12;0\right]
 \otimes[0]
 \;\oplus\;
 L\bar{L}
 \left[1+2\sqrt{7};\tfrac12,\tfrac12;0\right]
 \otimes[1]$
\\[2pt]

\hline

Gravitino
&
$\Big(
 L\bar{L}
 \left[\tfrac{13}{2};0,\tfrac12;1\right]
 \oplus
 L\bar{L}
 \left[\tfrac{13}{2};\tfrac12,0;-1\right]
 \Big)\otimes[1]$
\\[2pt]

\hline

Vector
&
$2L\bar{L}
 \left[1+\sqrt{37};0,0;0\right]
 \otimes[0]
 \;\oplus\;
 L\bar{L}
 \left[1+\sqrt{31};0,0;0\right]
 \otimes[1]$
\\[1pt]
&
$\oplus\;
 L\bar{L}
 \left[1+\sqrt{19};0,0;0\right]
 \otimes[2]$
\\[2pt]

\hline
\hline
& $k=1$ \\[2pt]
\hline

Graviton
&
$L\bar{L}
 \left[5;\tfrac12,\tfrac12;0\right]
 \otimes[0]$
\\[2pt]

\hline

Gravitino
&
$\Big(
 L\bar{A}_{1}
 \left[\tfrac92;0,\tfrac12;1\right]
 \oplus
 A_{1}\bar{L}
 \left[\tfrac92;\tfrac12,0;-1\right]
 \Big)\otimes[1]$
\\[2pt]

\hline

Vector
&
$L\bar{L}
 \left[1+\sqrt{19};0,0;0\right]
 \otimes[0]
 \;\oplus\;
 L\bar{L}
 \left[1+\sqrt{13};0,0;0\right]
 \otimes[1]$
\\[2pt]

\hline
\hline
& $k=0$ \\[2pt]
\hline

Graviton
&
$A_{1}\bar{A}_{1}
 \left[3;\tfrac12,\tfrac12;0\right]
 \otimes[0]$
\\[2pt]

\hline

Vector
&
$A_{2}\bar{A}_{2}
 \left[2;0,0;0\right]
 \otimes[1]
 \;\oplus\;
 L\bar{L}
 \left[1+\sqrt{7};0,0;0\right]
 \otimes[0]$
\\[2pt]

\hline
\hline

\end{tabular}%
}%
}

% Leave at least 12 percent of the text height for the caption.
\ifdim\dimexpr\ht0+\dp0\relax>0.88\textheight
  \resizebox*{!}{0.88\textheight}{\box0}%
\else
  \box0
\fi
\par
\endgroup

\caption{\footnotesize
The U$(1)_0$-invariant spectrum of $\mathrm{SU}(2,2|1)\times\mathrm{SU}(2)_{+}$ supermultiplets for MN1 at KK levels $k=0 , \ldots , 4$. Level $k=0$ reproduces the result
of \cite{Cassani:2020cod}.
}
\label{tab:MN1SinvariantMultiplets}
\end{table}

There is a tower of graviton multiplets, massless at KK level $k=0$ and long for $k \geq 1$, with quantum numbers:
\begin{equation} \label{MN1U1SinvGrav}
A_1 \bar{A}_1 \big[ 3 ; \tfrac12 , \tfrac12 ; 0 \big] \otimes [0] \; \oplus \;
\bigoplus_{k=1}^{\infty} \bigoplus_{\ell=0}^{m} L \bar{L}\big[ E^{(\textrm{grav})}_{k \ell}   ; \tfrac12 , \tfrac12 ; 0 \big] \otimes [\ell] \; .
\end{equation}
There is a tower of gravitino multiplets of both chiralities, starting at $k=1$ with short multiplets and continuing for $k \geq 2$ with long multiplets, with quantum numbers:
{\setlength\arraycolsep{2pt}
\begin{eqnarray} \label{MN1U1SinvGino}
& \Big( L \bar{A}_1 \big[ \tfrac92 ; 0 , \tfrac12 ; 1 \big] \oplus
A_1 \bar{L} \big[ \tfrac92 ; \tfrac12 , 0 ; -1 \big] \Big) \otimes [1] \nonumber \\[5pt]
& \oplus \; \bigoplus_{k=2}^{\infty} \bigoplus_{\ell=1}^{h}
\Big( L \bar{L} \big[ E^{(\textrm{gino})}_{k \ell} ; 0 , \tfrac12 ; 1 \big] \oplus
L \bar{L} \big[ E^{(\textrm{gino})}_{k \ell} ; \tfrac12 , 0 ; -1 \big] \Big) \otimes [\ell] \nonumber \\[5pt]
& \oplus \; \bigoplus_{k=2}^{\infty} \bigoplus_{\ell=1}^{h-1}
\Big( L \bar{L} \big[ E^{(\textrm{gino})}_{k \ell} ; 0 , \tfrac12 ; -1 \big] \oplus
L \bar{L} \big[ E^{(\textrm{gino})}_{k \ell} ; \tfrac12 , 0 ; 1 \big] \Big) \otimes [\ell] \; .
\end{eqnarray}
}Finally, the vector multiplet sector contains the massless flavour current along with a long multiplet at $k=0$. For all $k \geq 1$, there are towers of R-neutral long vectors. Starting at $k \geq 4$, there is a tower of long multiplets with superconformal primary R-charges $\pm2$. Specifically,
\begin{equation} \label{MN1U1SinvVec}
\begin{aligned}
& A_2 \bar{A}_2 \big[ 2 ; 0 , 0 ; 0 \big] \otimes [1]
\; \oplus \;
L \bar{L} \big[ 1+\sqrt{7} ; 0 , 0 ; 0 \big] \otimes [0]
\\[5pt]
& \oplus \; \bigoplus_{k=1}^{\infty} \bigoplus_{\ell=0}^{m+1}
\big( 2-\delta_{\ell,m}-\delta_{\ell,m+1} \big)  L \bar{L} \big[ E^{(\textrm{vec},0)}_{k\ell} ; 0 , 0 ; 0 \big] \otimes [\ell]
\\[5pt]
& \oplus \; \bigoplus_{k=1}^{\infty} \bigoplus_{\ell=2}^{m}
\Big( L \bar{L} \big[ E^{(\textrm{vec},2)}_{k\ell} ; 0 , 0 ; 2 \big]
\oplus L \bar{L} \big[ E^{(\textrm{vec},2)}_{k\ell} ; 0 , 0 ; -2 \big] \Big) \otimes [\ell] \; ,
\end{aligned}
\end{equation}
In (\ref{MN1U1SinvGrav})--(\ref{MN1U1SinvVec}) we have denoted a representation of $\textrm{SU}(2,2|1)\times\textrm{SU}(2)_+$ with the notation $X\bar{Y}[E; j_1 , j_2; r] \otimes [\ell]$. Here, $[\ell]$ stands for the $(2\ell+1)$-dimensional, spin-$\ell$ representation of SU$(2)_+$, and the $\textrm{SU}(2,2|1)$ supermultiplets $X\bar{Y}[E; j_1 , j_2; r]$ follow the conventions of section 2.2.1 of \cite{Cordova:2016emh}, with $j_{1\textrm{here}} = \frac12 j_{\textrm{there}}$, $j_{2\textrm{here}} = \frac12 \bar{\jmath}_{\textrm{there}}$, $n_{\textrm{here}} = r_{\textrm{there}}$. Except for the $k \geq 1$ R-neutral vector multiplets $L \bar{L} \big[ E^{(\textrm{vec},0)}_{k\ell} ; 0 , 0 ; 0 \big]$, which occur in (\ref{MN1U1SinvVec}) with the indicated degeneracy, all other multiplets in (\ref{MN1U1SinvGrav})--(\ref{MN1U1SinvVec}) are non-degenerate.

Unlike the MN1 putative spectrum described in appendix~\ref{sec:PutSpec}, the U$(1)_0$-invariant spectrum only contains supermultiplets in integer spin representations of the flavour group SU$(2)_+$. For easy reference, table \ref{tab:MN1SinvariantMultiplets} summarises the U$(1)_0$-invariant global spectrum up to KK level $k=4$. The globally defined spectrum further contains $\Sigma_2$-dependent Maass modes, see appendix~\ref{sec:GravOp}. As in \cite{Bhattacharya:2024tjw}, these are expected to also fill in long multiplets.

\subsection{Global MN2 spectrum} \label{sec:GlobalSpecMN2}

The MN2 spectrum presented in \cite{Bhattacharya:2024tjw} is what we are referring to in this paper as the putative spectrum. Extracting the U$(1)_1$-singlets following the process specified in section~\ref{sec:PutUniv} above allows us to find the global MN2 spectrum of $\textrm{SU}(2,2|2)$ supermultiplets. At KK level $k$, this is
\begin{equation} \label{eq:MN2GlobalSpectrum}
A_2\bar{A}_2[0;0]_{2k+2}^{(k;0)}
\;\oplus\;
B_1\bar{B}_1[0;0]_{2k+4}^{(k+2;0)}
\;\oplus\;
\bigoplus_{b=1}^{\lfloor k/2\rfloor}
L\bar{L}[0;0]_{D_{kb}}^{(k-2b;0)} \; ,
\end{equation}
where
\begin{equation} \label{eq:MN2GlobalSpectrumDim}
D_{kb}=\sqrt{4+2k(k+3)+2(k-2b)(k-2b+1)} \; .
\end{equation}
The  $\textrm{SU}(2,2|2)$ supermultiplets follow the notation of \cite{Bhattacharya:2024tjw} based on \cite{Cordova:2016emh}. In particular, the subindex corresponds to the superconformal dimension and the superindex summarises the $\textrm{SU}(2) \times \textrm{U}(1)$ R-charges: its first entry is the $\textrm{SU}(2)$ spin and its second entry is the $\textrm{U}(1)$ charge. The tower of long vector multiplets starts at $k=2$ with $L\bar{L}[0;0]_{2\sqrt6}^{(0;0)}$. The global spectrum also contains the separate $\Sigma_2$-dependent Maass modes discussed in \cite{Bhattacharya:2024tjw}. The Hall-Littlewood index match with the $\cN=2$ class $\cS$ field theory result of \cite{Gadde:2011uv} occurs within the global sector (\ref{eq:MN2GlobalSpectrum}).

%%%%%%%%%%%%%%%
%%%%%%%%%%%%%%%

\section{Discussion} \label{sec:Discussion}

%%%%%%%%%%%%%%%
%%%%%%%%%%%%%%%

Trombone gaugings of lower-dimensional supergravities typically signal an underlying reduction on a non-compact internal space. The maximally supersymmetric $D=5$ TCSO$(5,0,1;1)$-gauged supergravity is no exception: its generalised parallelisation is first realised on $B_2\times S^4$, where $B_2$ is the group manifold of the non-abelian two-dimensional Lie algebra and provides a group-manifold presentation of the hyperbolic plane. After implementing the topological twist through the appropriate E$_{6(6)}$ duality frame, the same local construction applies to the compact bundles $\Sigma_2\rtimes_{p,q}S^4$, with $\Sigma_2=H^2/\Gamma$, that arise near the AdS$_5$ throats of M5-branes wrapped on hyperbolic, unpunctured Riemann surfaces. The generic members of this family are the BBBW solutions, while the MN1 and MN2 geometries occur at the distinguished $p=q$ and $q=0$ endpoints, respectively. Thus, a single maximal five-dimensional theory provides a common local description of holographic backgrounds of class $\cS$ with $\cN=1$ supersymmetry, enhanced SU$(2)_+$ flavour symmetry at MN1 and enhanced $\cN=2$ supersymmetry at MN2.

A notable feature of our construction is that the ExGG implementation of the topological twist becomes, after flattening with the untwisted generalised frame, a constant E$_{6(6)}$ transformation of the five-dimensional embedding tensor. This explains why the direct and twisted internal geometries reduce to one and the same TCSO$(5,0,1;1)$ theory, with the twisting integers encoded in the duality frame. The known globally defined U$(1)_z$-invariant truncations are recovered as consistent subsectors of this maximal theory, and the explicit uplift formulae reproduce the generic BBBW and MN1 backgrounds.

We have then applied the trombone-augmented ExFT mass matrices of \cite{Pico:2026kch} to the MN1 vacuum. After interpreting the spectrum generated by the local maximal generalised frame as a putative reservoir and imposing invariance under the globally defined generalised U$(1)_0$-structure, we obtained the complete $\Sigma_2$-constant universal sector at arbitrary KK level. The resulting states assemble into infinite towers of $\textrm{SU}(2,2|1)\times\textrm{SU}(2)_+$ graviton, gravitino and vector multiplets. Their dimensions are controlled by a single closed expression, while their flavour representations and multiplicities are determined exactly. This all-level organisation is a consequence of the enhanced SU$(2)_+$ symmetry of MN1 and constitutes the principal spectral result of this paper.

A direct analysis of the eleven-dimensional graviton equation independently reproduces the graviton tower and makes the globality issue more precise. The U$(1)_0$ singlets are exactly the modes for which a coefficient constant on $\Sigma_2$ extends globally over every MN1 bundle. Non-singlet harmonics may nevertheless yield physical global modes when accompanied by sections of the appropriate automorphic line bundles; their masses are then controlled by weighted Maass eigenvalues. Thus, U$(1)_0$ invariance identifies a guaranteed universal $\Sigma_2$-constant sector, but it does not exhaust the spectrum associated with a specified compact quotient $\Sigma_2=H^2/\Gamma$.

From the dual SCFT perspective, our result supplies an all-level universal sector of the light single-trace operator spectrum of the class $\cS$ $\cN=1$ SCFT dual to MN1. The spectrum contains the stress-tensor and SU$(2)_+$ flavour-current multiplets, a protected pair of gravitino multiplets and infinite families of long graviton, gravitino and vector multiplets, with their R-charges, flavour representations and conformal dimensions determined holographically. Several extensions are natural. The same maximal truncation and mass-matrix framework can be applied to the generic BBBW vacua, where the dimensions vary with the twist parameter and the global spectrum exhibits a richer charged-sector structure \cite{Bhattacharya:2026iaa}. Also, for a fixed Riemann surface, it would be valuable to solve the relevant weighted Maass problems and determine how the additional $\Gamma$-dependent modes complete the universal multiplets found here.

%%%%%%%%%%%%%%%
%%%%%%%%%%%%%%%

\section*{Acknowledgements}

%%%%%%%%%%%%%%%
%%%%%%%%%%%%%%%

This work was supported by NSF grant PHY-2609785.

%%%%%%%%%%%%%%%
%%%%%%%%%%%%%%%

\appendix

\addtocontents{toc}{\setcounter{tocdepth}{1}}

%%%%%%%%%%%%%%%
%%%%%%%%%%%%%%%

%%%%%%%%%%%%%%%
%%%%%%%%%%%%%%%

\section{$\Ex{6}$ exceptional generalised geometry for M-theory}
\label{sec:Frames_EGGintro}

%%%%%%%%%%%%%%%
%%%%%%%%%%%%%%%

The bosonic fields of $D=11$ supergravity \cite{Cremmer:1978km} comprise the metric, $g_{11}$, a three-form potential, $A$, and a six-form potential, $\tilde{A}$. The latter are on-shell related by Hodge duality. The ExGG/ExFT \cite{Hohm:2013pua,Hohm:2013vpa,Coimbra:2011ky,Coimbra:2012af} approach to $D=11$ supergravity can be regarded as a rewrite of the latter that makes manifest the $\mathbb{R}^+ \times \mathrm{E}_{6(6)}$ symmetry of $D=5$ $\cN=8$ ungauged supergravity. This is achieved at the expense of foregoing the explicit local GL$(11)$ diffeomorphism covariance in favour of only manifest $\GLg{5}\times \GLg{6}$ covariance. This approach is, thus, ideal to describe $D=11$ supergravity on a direct product background $M_5 \times M_6$, where the manifolds $M_5$ and $M_6$ are respectively pseudo-Riemannian and Riemannian. 

This product structure is effected by splitting the $D=11$ spacetime coordinates as $(x^\mu , y^m)$, with $\mu =0, \ldots ,  4$ as in section \ref{sec:FieldContentET}, and $m=1, \ldots , 6$. The $D=11$ supergravity fields similarly break up into $\GLg{5}\times \GLg{6}$-covariant fields as 
\begin{equation} \label{eq:11DFields}
\begin{split}
g_{11}&= e^{2\tilde \Delta} g_{\mu \nu} d x^\mu d x^\nu + g_{mn} Dy^m Dy^n\,,\\[4pt]
A&= \tfrac{1}{3!} A_{mnp} Dy^m \wedge Dy^n \wedge Dy^p + \tfrac{1}{2!} A_{\mu np} d x^\mu \wedge Dy^n \wedge Dy^p + \dots\,,
\\[4pt]
\tilde A &= \tfrac{1}{6!} \tilde A_{m_1 \dots m_6} Dy^{m_1} \wedge \dots \wedge Dy^{m_6} + \tfrac{1}{5!}  \tilde A_{\mu m_1 \dots m_5} d x^\mu \wedge Dy^{m_1} \wedge \dots \wedge Dy^{m_5} + \dots\,,
\end{split}
\end{equation}
where $Dy^m \equiv d y^m + A_\mu{}^m dx^\mu$. All fields in (\ref{eq:11DFields}) depend on all $D=11$ coordinates,~{\it i.e.}~$g_{\mu\nu} = g_{\mu\nu} (x,y)$,~etc.,~and the dots stand for higher-degree GL$(5)$ forms whose explicit expression will not be needed. Now, the $\GLg{5}\times \GLg{6}$-covariant fields in (\ref{eq:11DFields}) can be equivalently regarded as components of the following manifestly $\GLg{5} \times \mathbb{R}^+ \times \mathrm{E}_{6(6)}$ covariant objects
\begin{eqnarray} \label{eq:ExObjects}
(G^{-1})^{MN} \Leftrightarrow \big( \tilde \Delta , g_{mn},A_{mnp}, \tilde{A}_{m_1 \dots m_6} \big) \; , \qquad 
\cA_\mu{}^M = \big( A_\mu{}^m, A_{\mu mn},  A_{\mu m n p q r}  \big) \, , \;
\end{eqnarray}
together with similar objects obtained out of the fields hidden under the ellipses in (\ref{eq:11DFields}). In particular, the $\GLg{5}$ one-form $\cA_\mu{}^M$ lies in the $\overline{\bm{27}}$ of $\mathrm{E}_{6(6)}$ because, under $\mathrm{E}_{6(6)} \supset \mathrm{GL}(6)$, 
\begin{equation} \label{eq:27Branching}
\overline{\bm{27}} \rightarrow \overline{\bm{6}} + \bm{15} + \bm{6}^\prime  \; ,
\end{equation}
corresponding to the three $\mathrm{GL}(6)$-covariant blocks shown for $\cA_\mu{}^M$ in (\ref{eq:ExObjects}). The expressions (\ref{eq:ExObjects}) are only schematic, as further non-linear redefinitions are needed, see \cite{Ciceri:2014wya}. Likewise, $(G^{-1})^{MN}$ is a $\GLg{5}$-covariant (inverse) metric on $\mathbb{R}^+ \times \mathrm{E}_{6(6)}/\mathrm{USp}(8)$. Accordingly, a generalised vielbein $E_M{}^{\uN}$, with inverse $\hat{E}^{M}{}_{\uN}$, can be defined as usual such that 
\begin{eqnarray} \label{eq:GenMet}
G_{MN} (x,y)= \delta_{\uP\uQ} E_M{}^{\uP} (x,y) E_N{}^{\uQ} (x,y) \; , \;
(G^{-1})^{MN} (x,y)= \delta^{\uP\uQ} \hat{E}^M{}_{\uP} (x,y) \hat{E}^N{}_{\uQ} (x,y) . \;
\end{eqnarray} 

The ExGG fields $G_{MN}(x,y)$, $\cA_\mu{}^M (x,y)$, etc., are covariant under E$_{6(6)}$ generalised diffeomorphisms, namely, under the combined action of GL(6) diffeomorphisms and gauge transformations of the component supergravity fields. The action of generalised diffeomorphisms is implemented with the help of the generalised Lie derivative, $L$. For any two generalised vectors $V^M = \big( v^m, \omega_{mn},  \sigma_{m n p q r}  \big)$, $V^{\prime M} = \big( v^{\prime m} , \omega^\prime_{mn},  \sigma^\prime_{m n p q r}  \big)$, this is defined as 
\begin{equation} \label{eq:LieDer}
L_V V^\prime \equiv  \big( \cL_v v^\prime \; , \; \cL_v \omega^\prime - \iota_{v^\prime} d\omega \; , \;  \cL_v \sigma^\prime- \iota_{v^\prime}  d\sigma - \omega^\prime \wedge d \omega \big)  \, ,
\end{equation}
in terms of the usual Lie derivative $\cL$, interior product $\iota$ and differential $d$ of standard geometry.

Finally, the ExGG E$_{6(6)}$-covariant fields may be interpreted as sections of suitable fibre bundles that extend the conventional tangent bundle of ordinary geometry. In this framework, one can naturally introduce generalised $G$-structures, defined as reductions to bundles with (compact) structure group $G \subset \textrm{USp}(8) \subset \textrm{E}_{6(6)}$. Such generalised $G$-structures are fully specified by $G$-invariant tensors together with an intrinsic torsion transforming in those representations of $G$ obtained from the branching of the $\bm{27}_{-1} + \overline{\bm{351}}_{-1}$ of $\mathbb{R}^+ \times \mathrm{E}_{6(6)}$ under $G \subset \textrm{USp}(8) \subset \textrm{E}_{6(6)}$. In particular, the generalised metric (\ref{eq:GenMet}) determines a generalised USp$(8)$-structure whose intrinsic torsion $X_{\uM \uN}{}^{\uP}(x,y)$ is defined via the generalised Lie derivative (\ref{eq:LieDer}) according to
\begin{equation} \label{eq:IntTor}
L_{\hat E_{\uM} (x,y)} \hat E_{\uN}(x,y) = - X_{\uM\uN}{}^{\uP} (x,y) \, \hat E_{\uP} (x,y)  \; .
\end{equation}
Here, we have again displayed the explicit $(x,y)$ dependence to emphasise that all quantities involved are eleven-dimensional.

%%%%%%%%%%%%%%%
%%%%%%%%%%%%%%%

\section{Geometric conventions and generalised identity structures}
\label{sec:GeomGISAppendix}

%%%%%%%%%%%%%%%
%%%%%%%%%%%%%%%

\subsection{$S^4$ conventions}
\label{sec:S4Conv}

%%%%%%%%%%%%%%%
%%%%%%%%%%%%%%%

We follow the notation and conventions of \cite{Cassani:2020cod} for the four-sphere $S^4$. The constrained coordinates $y^i$, $i=1, \ldots, 5$, that define $S^4$ as $\delta_{ij} y^i y^j =1$ are defined in terms of intrinsic $S^4$ angles $\zeta$, $\theta$, $\phi$, $\psi$ as
\begin{equation} \label{eq:S4angles}
y^1+i y^2 = e^{\frac{i}{2}(\phi+\psi)} \sin\zeta \cos\tfrac{\theta}{2} \; , \qquad 
y^3+i y^4 = e^{\frac{i}{2}(\phi-\psi)} \sin\zeta \sin\tfrac{\theta}{2} \; , \qquad 
y^5 = \cos\zeta  \; .
\end{equation}
In terms of these angles, the constrained coordinates $\mu_0$, $\mu_1$, $\mu_2$, with $\mu_0^2+\mu_1^2+\mu_2^2 =1$ and angles $\chi_1$, $\chi_2$ used to write the BBBW solution (\ref{eq:BBBWMetric}) are
\begin{equation} \label{eq:AltS4Angles}
 \mu_0=\cos\zeta \; ,  \quad
 \mu_1=\sin\zeta\cos\tfrac{\theta}{2}  \; ,  \quad
 \mu_2=\sin\zeta\sin\tfrac{\theta}{2}  \; ,  \quad
 \chi_1 = -\tfrac{\phi+\psi}{2} \; ,  \quad
 \chi_2 = \tfrac{\phi-\psi}{2} \; ,
\end{equation}
and the SU$(2)$ left-invariant one-forms that feature in the MN1 solution (\ref{eq:11DMetricMN1}) are
\begin{equation} \label{eq:SU2LIForms}
\sigma_1 = \cos\psi\, d\theta+ \sin\psi\,\sin\theta\,d\phi \ , \; 
\sigma_2 = -\sin\psi\, d\theta+ \cos\psi\,\sin\theta\,d\phi \ , \; 
\sigma_3 = d\psi+\cos\theta\,d\phi \; . 
\end{equation}

%%%%%%%%%%%%%%%
%%%%%%%%%%%%%%%

%%%%%%%%%%%%%%%
%%%%%%%%%%%%%%%

\subsection{Explicit GIS on $B_2\times S^4$ and its intrinsic torsion}
\label{sec:ExplicitGIS}

%%%%%%%%%%%%%%%
%%%%%%%%%%%%%%%

In this subsection we give the coordinate form of the generalised frame
\eqref{eq:GenFrameSplit2}, \eqref{eq:GenFrameB2S4}, compute its complete
generalised Lie derivative algebra and verify explicitly the constant-intrinsic-torsion
condition \eqref{eq:CTConst}.  We use the conventions of appendices
\ref{sec:Frames_EGGintro} and \ref{sec:S4Conv}.

\subsubsection{Geometric data and explicit frame}

On $B_2$, we use the coordinates, frame, volume form and metric given in
\eqref{eq:B2Geom}.  The corresponding inverse frame is
$\hat e_1=R_\Sigma^{-1}y\,\partial_x$ and
$\hat e_2=R_\Sigma^{-1}y\,\partial_y$.  It follows directly that
$de^1=R_\Sigma^{-1}e^1\wedge e^2$, $de^2=0$ and that $\hat e_1$, $\hat e_2$ satisfy the $B_2$ Lie algebra
\begin{equation} \label{eq:B2LieAlg}
[\hat e_1,\hat e_2]=-R_\Sigma^{-1}\hat e_1 \; .
\end{equation}
For the round four-sphere of radius $R$, the angular coordinates of
appendix~\ref{sec:S4Conv} give
\begin{equation}
\label{eq:ExplicitS4Metric}
ds^2(S^4)=R^2\left[
d\zeta^2+\frac14\sin^2\zeta\,
\left(\sigma_1^2+\sigma_2^2+\sigma_3^2\right)
\right]\;,\qquad
\vol_4=\frac{R^4}{8}\sin^3\zeta\,
d\zeta\wedge\sigma_{123}\;,
\end{equation}
where $\sigma_{123}\equiv\sigma_1\wedge\sigma_2\wedge\sigma_3
=\sin\theta\,d\theta\wedge d\phi\wedge d\psi$.
A convenient local gauge for the three-form potential is
\begin{equation}
\label{eq:ExplicitS4Potential}
A=\frac{R^3}{32}\,F(\zeta)\,\sigma_{123}\;,\qquad
F(\zeta)\equiv\cos(3\zeta)-9\cos\zeta\;,
\end{equation}
for which $dA=3R^{-1}\vol_4$.  With $\epsilon_{12345}=+1$, the Hodge
duals entering the frame are
\begin{equation}
\label{eq:ExplicitS4Hodge}
*_4dy_i
=-\frac{R^2}{3!}\epsilon_{ijklm}y^j
dy^k\wedge dy^l\wedge dy^m\;,\qquad
*_4(dy_i\wedge dy_j)
=\frac12\epsilon_{ijklm}y^kdy^l\wedge dy^m\;.
\end{equation}
The SO$(5)$ Killing vectors are normalised as
\begin{equation}
\label{eq:ExplicitS4Killing}
v_{ij}
=R^{-1}\left(
y_i\frac{\partial}{\partial y^j}
-y_j\frac{\partial}{\partial y^i}
\right)\Bigg|_{S^4}\;,\qquad
\iota_{v_{ij}}dy^k
=R^{-1}\left(y_i\delta_j{}^k-y_j\delta_i{}^k\right)\;.
\end{equation}
In particular, $v_{12}=R^{-1}(\partial_\phi+\partial_\psi)$ and
$v_{34}=R^{-1}(\partial_\phi-\partial_\psi)$.

It is useful to introduce the sphere forms
\begin{equation}
\label{eq:ExplicitSphereTensors}
Q_i\equiv-y_i\vol_4+R\,dy_i\wedge A\;,\qquad
P_i\equiv R*_4dy_i+y_iA\;,\qquad
W_{ij}\equiv R^2*_4(dy_i\wedge dy_j)+\iota_{v_{ij}}A\;.
\end{equation}
Equations \eqref{eq:ExplicitS4Metric}--\eqref{eq:ExplicitS4Hodge} yield
\begin{align}
\label{eq:ExplicitSphereTensorsExpanded}
Q_i
&=\frac{R^4}{32}
\left[
F(\zeta)\,\partial_\zeta y_i
-4y_i\sin^3\zeta
\right]d\zeta\wedge\sigma_{123}\;,
\nonumber\\[3pt]
P_i
&=-\frac{R^3}{3!}\epsilon_{ijklm}y^j
dy^k\wedge dy^l\wedge dy^m
+\frac{R^3}{32}F(\zeta)y_i\sigma_{123}\;,
\nonumber\\[3pt]
W_{ij}
&=\frac{R^2}{2}\epsilon_{ijklm}y^k
dy^l\wedge dy^m
+\frac{R^3}{32}F(\zeta)\,
\iota_{v_{ij}}\sigma_{123}\;.
\end{align}
Substitution into \eqref{eq:GenFrameB2S4} gives the following coordinate
expressions for the twenty-seven generalised vectors:
\begin{align}
\label{eq:ExplicitGISCoordinate}
\hat E_1
&=\left(R_\Sigma^{-1}y\,\partial_x,\;0,\;0\right)\;,\qquad
\hat E_2
=\left(R_\Sigma^{-1}y\,\partial_y,\;0,\;0\right)\;,
\nonumber\\[4pt]
\hat E^{1i}
&=\left(
0,\;
RR_\Sigma\frac{dx}{y}\wedge dy^i,\;
\frac{R_\Sigma R^4}{32}\frac{dx}{y}\wedge
\left[
F(\zeta)\partial_\zeta y^i-4y^i\sin^3\zeta
\right]d\zeta\wedge\sigma_{123}
\right)\;,
\nonumber\\[4pt]
\hat E^{2i}
&=\left(
0,\;
RR_\Sigma\frac{dy}{y}\wedge dy^i,\;
\frac{R_\Sigma R^4}{32}\frac{dy}{y}\wedge
\left[
F(\zeta)\partial_\zeta y^i-4y^i\sin^3\zeta
\right]d\zeta\wedge\sigma_{123}
\right)\;,
\nonumber\\[4pt]
\hat E_i
&=\left(
0,\;
R_\Sigma^2y_i\frac{dx\wedge dy}{y^2},\;
R_\Sigma^2\frac{dx\wedge dy}{y^2}\wedge
\left[
-\frac{R^3}{3!}\epsilon_{ijklm}y^j
dy^k\wedge dy^l\wedge dy^m
+\frac{R^3}{32}F(\zeta)y_i\sigma_{123}
\right]
\right)\;,
\nonumber\\[4pt]
\hat E_{ij}
&=\left(
v_{ij},\;
\frac{R^2}{2}\epsilon_{ijklm}y^kdy^l\wedge dy^m
+\frac{R^3}{32}F(\zeta)\iota_{v_{ij}}\sigma_{123},\;
0
\right)\; , 
\end{align}
where the functions $y^i(\zeta,\theta,\phi,\psi)$ are those in
\eqref{eq:S4angles}.

\subsubsection{Generalised Lie derivative algebra}

The $S^4$ quantities (\ref{eq:ExplicitSphereTensors}) satisfy the differential relations
\begin{equation}
\label{eq:ExplicitSphereDifferentials}
dQ_i=0\;,\qquad
dP_i=R^{-1}Q_i\;,\qquad
d\!\left(*_4dy_i\right)=-4R^{-2}y_i\vol_4\;.
\end{equation}
The second identity in \eqref{eq:ExplicitSphereDifferentials} fixes the
relative sign in $P_i=R*_4dy_i+y_iA$.
Using the definition \eqref{eq:LieDer} of the generalised Lie derivative, the $B_2$ relations (\ref{eq:B2LieAlg}), the Killing vector identities \eqref{eq:ExplicitS4Killing} and the differential relations 
\eqref{eq:ExplicitSphereDifferentials}, the complete set of non-vanishing
generalised Lie derivatives of the generalised frame (\ref{eq:ExplicitGISCoordinate}) w.r.t.~itself is
{\setlength\arraycolsep{3pt}
\begin{align}
\label{eq:ExplicitGISAlgebra}
L_{\hat E_1}\hat E_2
&=-R_\Sigma^{-1}\hat E_1\;,
&
L_{\hat E_2}\hat E_1
&=R_\Sigma^{-1}\hat E_1\;,
\nonumber\\[3pt]
L_{\hat E_1}\hat E^{1i}
&=R_\Sigma^{-1}\hat E^{2i}\;,
&
L_{\hat E_2}\hat E^{1i}
&=-R_\Sigma^{-1}\hat E^{1i}\;,
\nonumber\\[3pt]
L_{\hat E_2}\hat E_i
&=-R_\Sigma^{-1}\hat E_i\;,
\nonumber\\[5pt]
L_{\hat E^{1i}}\hat E_1
&=-R_\Sigma^{-1}\hat E^{2i}\;,
&
L_{\hat E^{1i}}\hat E_2
&=R_\Sigma^{-1}\hat E^{1i}\;,
\nonumber\\[3pt]
L_{\hat E^{1i}}\hat E_{jk}
&=2R_\Sigma^{-1}\delta^i{}_{[j}\hat E_{k]}\;,
\nonumber\\[5pt]
L_{\hat E_i}\hat E_1
&=-R^{-1}\hat E^{2i}\;,
&
L_{\hat E_i}\hat E_2
&=R^{-1}\hat E^{1i}\;,
\nonumber\\[3pt]
L_{\hat E_i}\hat E_{jk}
&=2R^{-1}\delta_{i[j}\hat E_{k]}\;,
\nonumber\\[5pt]
L_{\hat E_{ij}}\hat E^{xk}
&=-2R^{-1}\delta^k{}_{[i}\hat E^{xj]}\;,
&
L_{\hat E_{ij}}\hat E_k
&=-2R^{-1}\delta_{k[i}\hat E_{j]}\;,
\nonumber\\[3pt]
L_{\hat E_{ij}}\hat E_{kl}
&=-R^{-1}\left(
\delta_{ik}\hat E_{jl}
-\delta_{il}\hat E_{jk}
+\delta_{jl}\hat E_{ik}
-\delta_{jk}\hat E_{il}
\right)\;.
\end{align}
}All other generalised Lie derivatives not displayed in
\eqref{eq:ExplicitGISAlgebra} vanish.  Notice that the generalised Lie derivative
is not antisymmetric: for example,
$L_{\hat E_i}\hat E_1=-R^{-1}\hat E^{2i}$, whereas
$L_{\hat E_1}\hat E_i=0$.

%%%%%%%%%%
\subsubsection{Comparison with the embedding tensor}
%%%%%%%%%%

We finally compare \eqref{eq:ExplicitGISAlgebra} with the right-hand side
of \eqref{eq:CTConst}.  Substituting \eqref{eq:Charges1} into
\eqref{eq:XSymbols} and evaluating the resulting
$\mathbb{R}^+\times{\rm E}_{6(6)}$ generators on the flat basis
$\{\hat E_x,\hat E^{xi},\hat E_i,\hat E_{ij}\}$ gives the following
non-vanishing actions:
\begin{align}
\label{eq:ExplicitEmbeddingTensorAction}
-X_1\hat E_2
&=-g_2\hat E_1\;,
&
-X_1\hat E^{1i}
&=g_2\hat E^{2i}\;,
\nonumber\\[3pt]
-X_2\hat E_1
&=g_2\hat E_1\;,
&
-X_2\hat E^{1i}
&=-g_2\hat E^{1i}\;,
\nonumber\\[3pt]
-X_2\hat E_i
&=-g_2\hat E_i\;,
\nonumber\\[5pt]
-X^{1i}\hat E_1
&=-g_2\hat E^{2i}\;,
&
-X^{1i}\hat E_2
&=g_2\hat E^{1i}\;,
\nonumber\\[3pt]
-X^{1i}\hat E_{jk}
&=2g_2\delta^i{}_{[j}\hat E_{k]}\;,
\nonumber\\[5pt]
-X_i\hat E_1
&=-g_1\hat E^{2i}\;,
&
-X_i\hat E_2
&=g_1\hat E^{1i}\;,
\nonumber\\[3pt]
-X_i\hat E_{jk}
&=2g_1\delta_{i[j}\hat E_{k]}\;,
\nonumber\\[5pt]
-X_{ij}\hat E^{xk}
&=-2g_1\delta^k{}_{[i}\hat E^{xj]}\;,
\nonumber\\[3pt]
-X_{ij}\hat E_k
&=-2g_1\delta_{k[i}\hat E_{j]}\;,
\nonumber\\[3pt]
-X_{ij}\hat E_{kl}
&=-g_1\big(
\delta_{ik}\hat E_{jl}
-\delta_{il}\hat E_{jk}
+\delta_{jl}\hat E_{ik}
-\delta_{jk}\hat E_{il}
\big)\;.
\end{align}
All actions not displayed in \eqref{eq:ExplicitEmbeddingTensorAction}
vanish.  Using the coupling--radius relation \eqref{eq:CouplingsRadii},
these actions agree term by term with \eqref{eq:ExplicitGISAlgebra}
and therefore verify \eqref{eq:CTConst}.  The generalised frame
\eqref{eq:ExplicitGISCoordinate} therefore defines a GIS on
$B_2\times S^4$ whose constant intrinsic torsion is the embedding tensor
of $D=5$ $\cN=8$ TCSO$(5,0,1;1)$-gauged supergravity.

%%%%%%%%%%%%%%%
%%%%%%%%%%%%%%%

\subsection{Twisted GIS and its intrinsic torsion}
\label{sec:ExplicitTwistedGIS}

%%%%%%%%%%%%%%%
%%%%%%%%%%%%%%%

We now move on to give the finite flat-frame realisation of the twisted
GIS \eqref{eq:TwistedFrame} and verify the constant-intrinsic-torsion
condition \eqref{eq:CTConstTilde}.  We use throughout the untwisted frame
and its generalised Lie derivative algebra given in
appendix~\ref{sec:ExplicitGIS}.

\subsubsection{The tilded frame as a constant change of basis}

Introduce the antisymmetric matrix
\begin{equation}
\label{eq:TwistLambda}
\Lambda^i{}_{j}
=\frac{g_2}{g_1}\frac{1}{p+q}
\begin{pmatrix}
0&p&0&0&0\\
-p&0&0&0&0\\
0&0&0&-q&0\\
0&0&q&0&0\\
0&0&0&0&0
\end{pmatrix},
\end{equation}
and its Hodge dual in five dimensions,
\begin{equation}
\label{eq:TwistRho}
\rho_{ijk}\equiv\frac12\epsilon_{ijklm}\Lambda^{lm}.
\end{equation}
Its independent non-vanishing components are
$\Lambda^{12}=\frac{p}{p+q}\frac{g_2}{g_1}$,
$\Lambda^{34}=-\frac{q}{p+q}\frac{g_2}{g_1}$,
$\rho_{125}=\Lambda^{34}$ and $\rho_{345}=\Lambda^{12}$.
Exponentiating the constant Lie-algebra element
\eqref{eq:E7GroupElement}, or equivalently flattening and exponentiating
\eqref{eq:E6GroupElement} by means of \eqref{eq:FlatteningCon}, gives
{\setlength\arraycolsep{3pt}
\begin{align}
\label{eq:TwistedFrameFlatCombination}
\hat{\tilde E}_1
&=
\hat E_1
+\frac12\Lambda^{ij}\hat E_{ij}
-\frac18\epsilon_{ijklm}\Lambda^{ij}\Lambda^{kl}
 \hat E^{1m},
\nonumber\\[3pt]
\hat{\tilde E}_2
&=\hat E_2,
&
\hat{\tilde E}^{\,1i}
&=\hat E^{1i},
\nonumber\\[3pt]
\hat{\tilde E}^{\,2i}
&=\hat E^{2i}-\Lambda^i{}_{j}\hat E_j,
&
\hat{\tilde E}_i
&=\hat E_i,
\nonumber\\[3pt]
\hat{\tilde E}_{ij}
&=\hat E_{ij}-\rho_{ijk}\hat E^{1k}.
\end{align}
}The quadratic term in the first line is required by the finite
exponentiation; for the matrix \eqref{eq:TwistLambda}, it is supported
only along \(\hat E^{15}\).  Equation
\eqref{eq:TwistedFrameFlatCombination} is the flat-index form of the
coordinate-dependent transformation \eqref{eq:TwistedFrame}.  In
particular, its vector component gives
\(R_\Sigma^{-1}(y\partial_x+z\partial_\phi+\partial_\psi)\) in the
\(\hat{\tilde E}_1\) direction, consistently with
\eqref{eq:defz} and the Killing vectors displayed in
appendix~\ref{sec:ExplicitGIS}.

\subsubsection{Generalised Lie derivative and the transformed embedding tensor}

Let \(U_{\uM}{}^{\uN}\) denote the constant matrix defined by
\eqref{eq:TwistedFrameFlatCombination}.  Since its entries are constant,
bilinearity of the generalised Lie derivative and
\eqref{eq:CTConst} give
\begin{align}
\label{eq:TwistedGISConjugationCheck}
L_{\hat{\tilde E}_{\uM}}\hat{\tilde E}_{\uN}
&=
U_{\uM}{}^{\uQ}U_{\uN}{}^{\uR}
L_{\hat E_{\uQ}}\hat E_{\uR}
\nonumber\\
&=-
U_{\uM}{}^{\uQ}U_{\uN}{}^{\uR}
X_{\uQ\uR}{}^{\uS}
(U^{-1})_{\uS}{}^{\uP}
\hat{\tilde E}_{\uP}
\nonumber\\
&=-\tilde X_{\uM\uN}{}^{\uP}
\hat{\tilde E}_{\uP}.
\end{align}
The second line is exactly the duality transformation
\eqref{eq:TransXSymbol}.  It therefore proves
\eqref{eq:CTConstTilde} without repeating the complete algebra
\eqref{eq:ExplicitGISAlgebra} in the transformed basis. The complete transformed algebra is obtained from
\eqref{eq:ExplicitGISAlgebra} by the constant change of basis
\eqref{eq:TwistedFrameFlatCombination}.

It remains to identify the additional components generated by this
change of duality frame.  In the parametrisation
\eqref{eq:XSymbolsTilde}, they take the compact form
\begin{equation}
\label{eq:TwistedComponentsLambda}
\xi^{2ij}{}_{6}=-g_1\Lambda^{ij},
\qquad
\theta_{ij}{}^{6}{}_{5}=-6g_2\rho_{ij5},
\end{equation}
with the original components \eqref{eq:Charges1} unchanged.  Substitution
of \eqref{eq:TwistLambda} and \eqref{eq:TwistRho} into
\eqref{eq:TwistedComponentsLambda} reproduces precisely the four
components already displayed in \eqref{eq:ExtraCompsET}.  Thus, the finite
frame transformation
\eqref{eq:TwistedFrameFlatCombination}, the transformed embedding tensor
\eqref{eq:XSymbolsTilde} and the generalised Lie derivative condition
\eqref{eq:CTConstTilde} agree component by component.

\section{Kaluza--Klein trombone mass matrices} \label{sec:KKMassMat}

For reference, let us summarise the mass matrices that we employed in section \ref{sec:Spectra} to compute the putative MN1 KK spectrum. The relevant trombone-dependent KK mass matrices were first derived in \cite{Pico:2026kch}, building on \cite{Malek:2019eaz,Malek:2020yue,Varela:2020wty,Cesaro:2020soq}.  We henceforth fix $D=5$, $d=6$, $\alpha =6$, $\kappa=5$ in the expressions of \cite{Pico:2026kch}. 

The KK mass matrices depend on the $D=5$ $\cN=8$ embedding tensor, $X_{\uM \uN}{}^{\uP}$. It is convenient to extract the $\overline{\bm{351}}_{-1}$ and $\bm{27}_{-1}$ (trombone) components, $\Theta_{\uM \uN}{}^{\uP}$ and $\vartheta_{\uM}$, explicitly as \cite{deWit:2004nw,LeDiffon:2008sh}
\begin{eqnarray} \label{eq:defX}
X_{\uM \uN}{}^{\uP} \equiv
\Theta_{\uM \uN}{}^{\uP}
 +  \Big(  \tfrac{9}{2} \,  \mathbb{P}^{\uQ}{}_{\uM}{}^{\uP}{}_{\uN}  - \delta_{\uM}^{\uQ}\delta_{\uN}^{\uP} \Big)
\vartheta_{\uQ}  \; , \quad 
\textrm{with \, $\Theta_{\uM \uN}{}^{\uP} \equiv \Theta_{\uM}{}^\alpha (t_\alpha)_{\uN}{}^{\uP}$} \; .
\end{eqnarray}
Equivalently, the $\overline{\bm{351}}_{-1}$ components $\Theta_{\uM \uN}{}^{\uP}$ can be repackaged in an antisymmetric matrix $Z^{\uM \uN}$ defined through either expression \cite{deWit:2004nw,LeDiffon:2008sh}
\begin{equation} \label{eq:ZembTen}
Z^{\uM \uN} = \Theta_{\uP \uQ}{}^{\uM} \, d^{\uN \uP \uQ}  \; , \qquad 
Z^{\uM \uN} = 2 \, \Theta_{\uP \uQ}{}^{\uT} \, d^{\uP \uR \uM} \, d^{\uQ \uS \uN} \, d_{\uR \uS \uT} \; .
\end{equation}
In (\ref{eq:defX}), $(t_\alpha)_{\uM}{}^{\uN}$ are the $\bm{78}$ generators of E$_{6(6)}$ in the $\bm{27}$-dimensional representation (see appendix A of \cite{Varela:2025xeb} for our conventions), and $\mathbb{P}^{\uK}{}_{\uM}{}^{\uL}{}_{\uN} \equiv (t^\alpha)_{\uM}{}^{\uK} (t_\alpha)_{\uN}{}^{\uL}$ the projector to the $\bm{78}$ of E$_{6(6)}$. The adjoint index $\alpha =1 , \ldots , 78$ is raised here with the (inverse of the) Cartan-Killing form, $\kappa_{\alpha \beta} \equiv \mathrm{tr} ( t_{\alpha}t_{\beta} )$. In (\ref{eq:ZembTen}), $d_{\uM \uN \uP}$ and $d^{\uM \uN \uP}$ denote the E$_{6(6)}$ cubic invariant and its conjugate. The mass matrices also depend on the generators, $(\cT_{\uM})^\Sigma{}_\Lambda$, of SO(5) in the $\oplus_{k=0}^{\infty} [k0]$ representation, as in the main text. These generators are subject to the relations (\ref{eq:LieAlgHarm}). 

With these conventions, we simply import the bosonic KK mass matrices from \cite{Pico:2026kch}, except for the tensor mass matrix (\ref{eq:KKTensor}), which is new. The KK graviton mass matrix is
\begin{equation} \label{eq:KKgraviton}
(\cM^2_{\textrm{graviton}})^\Sigma{}_\Lambda = - R^{-2} M^{ \uM \uN } \big( \fT_{\uM}{} \fT_{\uN} \big)^\Sigma{}_\Lambda -3R^{-1} M^{\uM\uN}  \vartheta_{\uM} (\fT_{\uN})^\Sigma{}_\Lambda \; ,
\end{equation}
the KK gauge field mass matrix reads
{\setlength\arraycolsep{2pt}
\begin{eqnarray} \label{eq:KKvector}
(\cM^2_{\textrm{vector}})^{\uM\Sigma}{}_{\uN\Lambda} & = & M^{\uM\uA'} \bigg[  \tfrac{1}{3} \left( X_{\uA'\uC}{}^{\uD} - 18 \,  \delta_{\uA'}^{\uD} \vartheta_{\uC} \right) M^{\uC \uE} \left( X_{\uN(\uD}{}^{\uF} M_{\uE)\uF} + \vartheta_{\uN} M_{\uD \uE} \right) \delta_\Lambda^\Sigma \nonumber  \\[5pt]
&& \qquad \quad \; +2 R^{-1} \left( X_{\uA' \uC}{}^{\uD} - 18 \, \delta_{\uA'}^{\uD} \,  \vartheta_{\uC} \right) M^{\uC \uE}  M_{\uF( \uD} \mathbb{P}^{\uF}{}_{\uE)}{}^{\uG}{}_{\uN} \,  ( \fT_{\uG})^\Sigma{}_\Lambda \nonumber  \\[5pt] 
&& \qquad  \quad \;   - 2 R^{-1} M^{\uC \uD} \left(  X_{\uN(\uA'}{}^{\uE} M_{\uD) \uE} + \vartheta_{\uN} M_{\uA' \uD} \right) ( \fT_{\uC})^\Sigma{}_\Lambda \nonumber  \\[5pt] 
&&  \qquad   \quad \;  - 12 R^{-2}  M^{\uC \uD}  M_{\uE(\uA'} \mathbb{P}^{\uE}{}_{\uD)}{}^{\uF}{}_{\uN}  ( \fT_{\uC} \fT_{\uF})^\Sigma{}_\Lambda  \bigg] \; ,
\end{eqnarray}
}the KK two-form mass matrix takes the form
{\setlength\arraycolsep{2pt}
\begin{eqnarray} \label{eq:KKTensor}
(\cM_{\textrm{tensor}})^{\uN \Sigma}{}_{\uM \Lambda} & = & \sqrt{\tfrac25} \, M_{\uM \uP} \,  \Big( \big( Z^{\uP \uN} -\tfrac{15}{2}  \,  d^{\uP \uN \uQ}   \, \vartheta_{\uQ}  \big) \delta^\Sigma_\Lambda + 5 \,  d^{\uP \uN \uQ}   \, ( \fT_{\uQ})^\Sigma{}_\Lambda  \Big)  \; ,
\end{eqnarray}
}and, finally, the KK scalar mass matrix reads
{\setlength\arraycolsep{1pt}
\begin{eqnarray} \label{eq:KKscalar}
%\begin{split}
 (\cM^2_{\textrm{scalar}})^{\alpha\Sigma}{}_{\beta\Lambda} &=&
\left( P_{\textrm{coset}} \right)^{\alpha \uA\uB} \Big[ \Theta_{\uA\uE}{}^{\uF} \Theta_{\uC\uF}{}^{\uE} M_{\uB\uD} \, \delta_\Lambda^\Sigma \nonumber  \\[4pt]
&& +\tfrac{1}{5} M_{\uB\uD} \Big( M_{\uF\uF'} M^{\uE\uE'}\Theta_{\uA\uE}{}^{\uF} \Theta_{\uC\uE'}{}^{\uF'}  + M_{\uF\uF'} M^{\uE\uE'}\Theta_{\uE\uA}{}^{\uF} \Theta_{\uE'\uC}{}^{\uF'} \nonumber   \\
&& \qquad  \qquad \quad + M_{\uA\uA'} M_{\uC\uC'} M^{\uF\uF'} M^{\uE\uE'}\Theta_{\uE\uF}{}^{\uA'} \Theta_{\uE'\uF'}{}^{\uC'} \Big) \delta_\Lambda^\Sigma \nonumber  \\[4pt]
&&   +\tfrac{2}{5} \Big(  M_{\uE\uF}\Theta_{\uA\uC}{}^{\uE} \Theta_{\uB\uD}{}^{\uF} - M_{\uC\uC'} M_{\uD\uD'} M^{\uE\uF}\Theta_{\uA\uE}{}^{\uC'} \Theta_{\uB\uF}{}^{\uD'} \nonumber  \\
&& \qquad  \qquad \quad - M_{\uC\uC'} M_{\uD\uD'} M^{\uE\uF}\Theta_{\uE\uA}{}^{\uC'} \Theta_{\uF\uB}{}^{\uD'} \Big) \delta_\Lambda^\Sigma \nonumber \\[4pt]
&& + 6 \Big( \vartheta_{\uC}  \Theta_{\uD\uA}{}^{\uB'} M_{\uB'\uB} - \vartheta_{\uE} \Theta_{\uF\uA}{}^{\uC'}  M_{\uB\uD}  M^{\uE\uF}  M_{\uC'\uC} \Big) \delta_\Lambda^\Sigma 
\nonumber  \\[4pt]
&& -2R^{-1} M_{\uE\uD} \Theta_{\uA\uC}{}^{\uE} (\mathcal{T}_{\uB})^\Sigma{}_\Lambda  +2R^{-1} M_{\uE\uB}  \Theta_{\uC\uA}{}^{\uE}  ( \mathcal{T}_{\uD})^\Sigma{}_\Lambda  \nonumber  \\[4pt]
&&  -4R^{-1} M^{\uE\uF} M_{\uB\uD} M_{\uC\uC'} \Theta_{\uF\uA}{}^{\uC'}    (\mathcal{T}_{\uE})^\Sigma{}_\Lambda  \nonumber  \\[4pt]
&&  +36 R^{-1}  M_{\uB\uD} \vartheta_{\uC}  (\mathcal{T}_{\uA})^\Sigma{}_\Lambda   +2\alpha R^{-2} M_{\uB\uD}   (\mathcal{T}_{\uC} \mathcal{T}_{\uA})^\Sigma{}_\Lambda  \Big] \left( P_{\textrm{coset}}  \right)_\beta{}^{\uC\uD}  \nonumber  \\[4pt]
&&  -\delta^\alpha_\beta \,  \big[ R^{-1} (D-2)  M^{\uA\uB} \vartheta_{\uA} (\fT_{\uB})^\Sigma{}_\Lambda + R^{-2} M^{\uA\uB} ( \fT_{\uA} \fT_{\uB})^\Sigma{}_\Lambda \big] \; . 
%
%\end{split}
\end{eqnarray}
}In the latter equation we have used the projector to the $\textrm{E}_{6(6)}/\textrm{USp}(8)$ coset, defined as \cite{Berman:2019izh}
\begin{equation} \label{eq:ProjCoset}
\phi^{\uM\uN \Lambda} = (P_{\textrm{coset}})_\alpha{}^{\uM\uN} \, \phi^{\alpha  \Lambda} \; . \qquad 
\textrm{with} \quad 
(P_{\textrm{coset}})_\alpha{}^{\uM\uN} \equiv (t_\alpha)^{(\uM}{}_{\uP} M^{\uN)\uP} \; .
\end{equation}

In all of the above expressions, the scalar matrix $M_{\uM\uN}$ and its inverse $M^{\uM\uN}$ must be evaluated at a vacuum of $D =5$ $\cN=8$ supergravity. As usual, the eigenvalues of (\ref{eq:KKgraviton}), (\ref{eq:KKvector}), (\ref{eq:KKscalar}) correspond to square masses, as indicated by the superscript. The tensor mass eigenvalues of (\ref{eq:KKTensor}) are linear masses. The spectrum following from \eqref{eq:KKgraviton} consists entirely of physical modes, which are identified with spin-two states. The situation is different for the rest of the bosonic mass matrices, whose spectra also include unphysical entries. At each KK level $k$, the vector and tensor mass matrices (\ref{eq:KKvector}), (\ref{eq:KKTensor}) respectively contain only $17 \, \textrm{dim} [k0]$ and $10 \, \textrm{dim} [k0]$ non-zero eigenvalues. All of these are physical in the two-form sector, but the vector mass matrix contains Goldstone modes of normalised squared mass $L^2 M^2_{1 \, \textrm{Goldstone}}$ related to a physical graviton mass $L^2 M_2^2$ by 
\begin{equation} \label{eq:HiggsingVec}
L^2 M^2_{1 \, \textrm{Goldstone}}= \tfrac83 \, L^2 M_2^2 +8 \; .
\end{equation}
Analogous Goldstone modes also occur in the scalar sector \eqref{eq:KKscalar}, realising the Higgs mechanism for both graviton and vector states.

Finally, the fermionic KK mass matrices can be written in terms of the $D=5$ fermion shifts $A_1^{ij}$, $A_2^{i,jkl}$ and $B^{ij}$ \cite{deWit:2004nw,LeDiffon:2008sh,Varela:2025xeb}, which are respectively related to the $\overline{\bm{351}}_{-1}$ and the $\bm{27}_{-1}$, trombone, components of the embedding tensor. The KK gravitino mass matrix reads
\begin{equation} \label{eq:KKgravitino}
(\cM_{\textrm{gravitino}})^{ij\,\Sigma}{}_{\Lambda}
=
\tfrac{3}{\sqrt{2}}\left[
\left(A_1^{ij}+2B^{ij}\right)\delta^\Sigma_\Lambda
-\tfrac{4}{3}R^{-1}(\cV^{-1})^{ij\uM}
(\fT_{\uM})^\Sigma{}_{\Lambda}
\right] \; ,
\end{equation}
while the KK spin-$1/2$ fermion mass matrix is
{\setlength\arraycolsep{0pt}
\begin{eqnarray} \label{eq:KKspin12}
(\cM_{\textrm{spin-$\frac12$}})^{ijk\,\Sigma}{}_{\ell mn\,\Lambda}
&=&9\sqrt{2}\Big[ \Big(
4\,\delta^{[[i}_{[[\ell}\Omega_{m|p|}\Omega_{n]]q}
A_2^{j,k]]pq}
+\delta^{[[i}_{[[\ell}\delta^j_m\Omega_{n]]p}
A_1^{k]]p} 
+6\,\delta^{[[i}_{[[\ell}\delta^j_m\Omega_{n]]p}
B^{k]]p} \Big) \delta^\Sigma_\Lambda \nonumber \\[4pt]
&&\qquad +4R^{-1}\,\delta^{[[i}_{[[\ell}\delta^j_m\Omega_{n]]p}
(\cV^{-1})^{k]]p\uM}
(\fT_{\uM})^\Sigma{}_{\Lambda}
\Big] \; .
\end{eqnarray}
}In these expressions, we have restored USp$(8)$ indices $i=1, \ldots , 8$, and have used double brackets to denote symplectic antisymmetrisation. For $(\fT_{\uM})^\Sigma{}_{\Lambda}=0$, and ignoring KK indices, (\ref{eq:KKgravitino}) and (\ref{eq:KKspin12}) reduce to the fermionic mass matrices, (2.33) and (2.34) of \cite{Varela:2025xeb}, of $D=5$ $\cN=8$ trombone-gauged supergravity. The eigenvalues of both matrices correspond to linear masses. All modes arising from (\ref{eq:KKgravitino}) are physical, while (\ref{eq:KKspin12}) also contains Goldstino modes of normalised eigenvalue $ L M_{\frac12  \, \textrm{Goldstino}}$ eaten by a massive gravitino of physical mass $L M_{\frac32}$, related through
\begin{equation} \label{eq:HiggsingFer}
 L M_{\frac12  \, \textrm{Goldstino}} =  \tfrac53 \, L M_{\frac32} \; .
\end{equation}
%

%%%%%%%%%%%%%%%
%%%%%%%%%%%%%%%

\section{Direct graviton analysis and the global MN1 sectors} \label{sec:GravOp}

%%%%%%%%%%%%%%%
%%%%%%%%%%%%%%%

In this appendix we derive the MN1 graviton spectrum directly from the eleven-dimensional fluctuation equation, independently of the ExFT mass-matrix calculation. Besides reproducing the U$(1)_0$-invariant tower, this analysis clarifies the relation between global definiteness and weighted Maass eigensections on $\Sigma_2$. 

The allowed graviton squared masses $(ML)^2$, normalised to the AdS$_5$ radius $L$ in (\ref{eq:MN1vac}), are determined by the eigenvalue equation \cite{Bachas:2011xa}
\begin{equation} \label{eq:MN1_BE1}
\frac{e^{-9\Delta}}{\sqrt{\text{det} \, g_6}}\partial_m\left(e^{9\Delta}\sqrt{\text{det} \, g_6}\ g_6^{mn}\partial_n h(y)\right) =-M^2L^2 \, h(y) 
\end{equation} 
with eigenfunction $h(y)$. Here, $g_6$ is the internal six-dimensional metric in (\ref{eq:11DMetricMN1}), and $e^{2\Delta}=\left(\frac{2}{3}\right)^{2/3}(3+\cos^2\zeta)^{1/3}$ the warp factor. We collectively denote the six-dimensional coordinates with $y \equiv (x,y, \zeta , \theta, \phi ,\psi)$, where $x,y$ are the coordinates on $\Sigma_2$ used in (\ref{eq:B2Geom}), and $\zeta$, $\theta$, $\phi$, $\psi$ the $S^4$ coordinates used in appendix \ref{sec:S4Conv}. Note the double use of $y$, which should not cause confusion. Now, (\ref{eq:MN1_BE1}) can be brought to the form
\begin{equation} \label{eq:MN1_diffeq}
\left[\partial_\zeta^2+3\cot\zeta\,\partial_\zeta
+g_\Sigma^{ab}(\nabla_a-\upsilon_a\partial_\psi)
 (\nabla_b-\upsilon_b\partial_\psi)
+\frac{3+\cos^2\zeta}{\sin^2\zeta}\,\Box_3\right] h(y) =-\tfrac13  (ML)^2 \, h(y) \ .
\end{equation}
where $g_{\Sigma}^{ab}$, $a=1,2$, is the inverse metric on $\Sigma_2$ defined in (\ref{eq:B2Geom}), $\nabla_a$ the covariant derivative on $\Sigma_2$, $\upsilon_a$ the components of the spin connection (\ref{eq:SpinCon}) on $\Sigma_2$, and $\Box_3$ denotes the Laplacian corresponding to the round $S^3$ associated to the SU$(2)_+$ left-invariant one-forms (\ref{eq:SU2LIForms}). 

We now propose the following factorised form for the eigenfunction $h(y)$,
\begin{equation} \label{eq:MN1_sepvar}
h(y)  = \varphi(x,y) \, {\cal Y} (\zeta, \theta, \phi, \psi) = \varphi(x,y) \, w^{k_1}(1-w)^{\ell}H(w) Y_{\ell m}^{(n/2)} (\theta,\phi)e^{i\frac{n}{2}\psi}  \; ,
\end{equation} 
where $\varphi(x,y)$ is a function on $\Sigma_2$, and ${\cal Y}$ after the first equality stands for the $S^4$ spherical harmonics (\ref{eq:harmS4}). A convenient parametrisation for the latter is introduced after the second equality in (\ref{eq:MN1_sepvar}), with $k_1$ a constant and $H(w)$ a function of $w \equiv \cos^2\zeta$, both of them to be fixed, and $Y_{\ell m}^{(n/2)} (\theta,\phi)$ the monopole harmonics on the $S^2$ base of the $S^3$ inside $S^4$. Note that the $S^3$ spherical harmonics are $\Phi_{\ell m n}(\theta,\phi,\psi)
 \equiv Y^{(n/2)}_{\ell m}(\theta,\phi) e^{\frac{i n}{2}\psi} $ and obey 
\begin{equation} \label{eq:MN1S3Eig}
\Box_3\Phi=-\frac{r(r+2)}{4}\,\Phi=-\ell(\ell+1)\Phi\, ,
\qquad r\equiv 2\ell\, .
\end{equation}
The quantum numbers $\ell$, $m$, $n$ in (\ref{eq:MN1_sepvar}) thus range as
\begin{equation} \label{eq:MN1HarmRanges}
\ell=0,\tfrac12,1,\tfrac32,\ldots\, ,
\qquad m=-\ell,-\ell+1,\ldots,\ell\, ,
\end{equation}
and
\begin{equation} \label{eq:MN1WeightSet}
 n\in W_r\equiv\{r,r-2,\ldots,-(r-2),-r\}\, ,
 \qquad r=2\ell\, .
\end{equation}

In order to ensure regularity of the eigenfunction, and with the benefit of hindsight, the constant $k_1$ in (\ref{eq:MN1_sepvar}) can be set to either of two values, $k_1=0$ or $k_1 = \tfrac12$. Let us start by setting $k_1 =0$. Under the factorised form  (\ref{eq:MN1_sepvar}), equation (\ref{eq:MN1_BE1}) becomes
\begin{eqnarray} \label{eq:MN1_hypg1}
& \left[4w(1-w)H''(w)+2(1-5w-4\ell w)H'(w)+\left(\frac{M^2L^2}{3}-\frac{n^2}{4}-\ell(5+3\ell)\right)H(w)\right]\varphi(x,y) \nonumber \\[5pt]
& +H(w) \, D_{n/2} \, \varphi (x,y) =0 \; , 
\end{eqnarray}
where we have introduced the Maass Laplacian of weight $\frac{n}{2}$,
\begin{equation} \label{eq:MaassLap}
D_{n/2} \equiv y^2\left(\frac{\partial^2}{\partial x^2}+\frac{\partial^2}{\partial y^2}\right)+iny\frac{\partial}{\partial x} \; .
\end{equation}
Thus, the differential equation (\ref{eq:MN1_hypg1}) separates into the Maass eigenvalue problem,
\begin{equation} \label{eq:MN1MaassEig}
D_{n/2} \, \varphi (x,y) =E_{n/2} \,\varphi (x,y)  \, ,
\end{equation}
with the $n$-dependent eigenvalue $E_{n/2}$ fulfilling the role of separation constant, and an equation for $H(w)$. The latter can be brought to hypergeometric form,
\begin{equation} \label{eq:HGEq}
w(1-w)H''(w)+(c-(a+b+1)w)H'(w)-abH(w)=0 \; ,
\end{equation}
with constants
\begin{eqnarray} \label{eq:HGEqConst}
a&=&\tfrac{3}{4}+\ell-\sqrt{\tfrac{1}{4}\left(\tfrac{1}{3} M^2L^2 +E_{n/2}-\tfrac{1}{4}n^2 -\ell(5+3\ell)\right)+\left(\tfrac{3}{4}+\ell\right)^2},\nonumber\\[5pt]
b&=&\tfrac{3}{4}+\ell+\sqrt{\tfrac{1}{4}\left(\tfrac{1}{3} M^2L^2 +E_{n/2}-\tfrac{1}{4}n^2 -\ell(5+3\ell)\right)+\left(\tfrac{3}{4}+\ell\right)^2}, \nonumber \\[5pt]
c&=&\tfrac{1}{2} \; .
\end{eqnarray}

Now, regularity at both endpoints $w=0$, $w=1$, demands that the hypergeometric function $H(w)$ actually be a polynomial, in agreement with the definition (\ref{eq:MN1_sepvar}) in terms of $S^4$ spherical harmonics (\ref{eq:harmS4}). This requires that $a$ in (\ref{eq:HGEqConst}) takes on negative integer values, $a= -\tilde{j}$, for $\tilde{j} = 0 , 1 , 2 , \ldots$ and, thus, imposes a quantisation condition on the graviton masses,
\begin{equation} \label{eq:QuantGravMass}
M^2L^2=3k(k+3)+\tfrac{3}{4} n^2 -3\ell(\ell+1)-3E_{n/2} \ ,
\end{equation}
where we have defined $k \equiv 2\tilde{j}+2\ell$. A similar analysis with $k_1  = \tfrac12$ in (\ref{eq:MN1_sepvar}) leads to the same condition (\ref{eq:QuantGravMass}) with, now, $k$ defined as $k \equiv 2\tilde{j}+2\ell +1 $. The MN1 graviton masses are thus (\ref{eq:QuantGravMass}) with $k$ any non-negative integer $k=0 ,1 , 2 , \ldots$

Let us finally spell out the relation between the Maass problem (\ref{eq:MN1MaassEig}), U$(1)_0$ invariance and global definiteness. The identification of the U$(1)_0$ charge with the integer $n$ in (\ref{eq:MN1_sepvar}) is specific to the graviton tower and follows from two observations. First, the embedding (\ref{eq:branching}) gives
\begin{equation} \label{eq:MN1U1zCharge}
\textrm{U}(1)_z=z\,\textrm{U}(1)_+ +\textrm{U}(1)_-+\textrm{U}(1)\, .
\end{equation}
The five-dimensional graviton is a USp$(8)$ singlet and therefore carries no charge under the last U$(1)$ factor in (\ref{eq:MN1U1zCharge}). Its U$(1)_z$ charge is consequently entirely inherited from the $S^4$ harmonic in $[k0]$. Second, at the BBBW vacuum the angular covariant derivatives (\ref{eq:AngCovDer}) reduce schematically to $D\phi=d\phi-zA$ and $D\psi=d\psi-A$. Thus, up to an overall sign convention, the action of U$(1)_z$ on a scalar $S^4$ harmonic is generated by
\begin{equation} \label{eq:MN1U1zGenerator}
K_z=z\,\partial_\phi+\partial_\psi\, .
\end{equation}
With $-i\partial_\phi Y_{\ell m}^{(n/2)}=mY_{\ell m}^{(n/2)}$ and the explicit $e^{in\psi/2}$ dependence in (\ref{eq:MN1_sepvar}), the harmonic has charge
\begin{equation} \label{eq:MN1U1zHarmCharge}
Q_z=z\,m+\frac{n}{2}\, .
\end{equation}
At the MN1 point, $z=0$, this reduces to $Q_0=n/2$. Hence, for the graviton tower,
\begin{equation} \label{eq:MN1U10Singlet}
\textrm{U}(1)_0\textrm{-invariance}\qquad\Longleftrightarrow\qquad n=0\, .
\end{equation}
The quantum number $m$ remains unrestricted, so the surviving modes fill a complete spin-$\ell$ representation of the unbroken SU$(2)_+$ symmetry. This conclusion must not be applied unchanged to the other spin towers. If the five-dimensional seed field transforms in a non-trivial USp$(8)$ representation $\bm r$, its intrinsic charge may cancel the charge of the harmonic, and the correct prescription is to select the U$(1)_0$ singlets in the full product $\bm r\otimes[k0]$, as in (\ref{eq:branchingKK}).

The same result can be seen directly from patching. On the overlap of two coordinate patches of $\Sigma_2$, the spin connection and the fibred angle transform as
\begin{equation} \label{eq:MN1Patching}
\upsilon_{(\alpha)}=\upsilon_{(\beta)}+d\lambda_{\alpha\beta}\, ,
\qquad
\psi_{(\alpha)}=\psi_{(\beta)}-\lambda_{\alpha\beta}\, .
\end{equation}
The charged factor in (\ref{eq:MN1_sepvar}) therefore obeys
\begin{equation} \label{eq:MN1HarmPatching}
e^{\frac{i n}{2}\psi_{(\alpha)}}
=e^{-\frac{i n}{2}\lambda_{\alpha\beta}}
 e^{\frac{i n}{2}\psi_{(\beta)}}\, .
\end{equation}
For $n\neq0$, a globally defined fluctuation can still be formed provided that $\varphi$ is a section of the oppositely charged automorphic line bundle,
\begin{equation} \label{eq:MN1CoefficientPatching}
\varphi_{(\alpha)}
=e^{\frac{i n}{2}\lambda_{\alpha\beta}}\varphi_{(\beta)}\, ,
\qquad
\varphi_{(\alpha)}e^{\frac{i n}{2}\psi_{(\alpha)}}
=\varphi_{(\beta)}e^{\frac{i n}{2}\psi_{(\beta)}}\, .
\end{equation}
A coefficient that is constant in a local trivialisation does not have this transition law unless $n=0$. Thus, among the putative graviton modes obtained by taking $\varphi$ to be constant on $\Sigma_2$, precisely the U$(1)_0$ singlets extend globally over every MN1 bundle. For these modes, $E_0=0$, $n=0$ and $\ell$ is integral. Equation (\ref{eq:QuantGravMass}) gives
\begin{equation} \label{eq:MN1GlobalGravMassMultiplicity}
M^2L^2=3k(k+3)-3\ell(\ell+1)\, ,
\qquad
\mathfrak d_{k,\ell}=2\ell+1\, ,
\qquad
\ell=0,1,\ldots,\left\lfloor\frac{k}{2}\right\rfloor\, .
\end{equation}
The multiplicity $\mathfrak d_{k,\ell}$ is the number of allowed values $m=-\ell,-\ell+1,\ldots,\ell$ and therefore is also the dimension of the corresponding SU$(2)_+$ representation.

The U$(1)_0$-singlet sector is a guaranteed global subsector, but it need not exhaust the global spectrum on a specified Riemann surface. For $n\neq0$, the allowed global coefficients are determined by the Maass eigenvalue problem (\ref{eq:MN1MaassEig}) on the relevant automorphic line bundle. For even $n\neq0$, so that the weight $n/2$ is integral, the universal discrete-series eigenvalues are \cite{Elstrodt:1973,Bhattacharya:2024tjw}
\begin{equation} \label{eq:MN1MaassUniv}
E^{\mathrm{univ}}_{n/2,j}
=\left(\frac{|n|}{2}-j\right)
 \left(\frac{|n|}{2}-j-1\right)\, ,
\qquad
j=0,1,\ldots,\frac{|n|}{2}-1\, ,
\end{equation}
with multiplicity \cite{Shimura:1971}
\begin{equation} \label{eq:MultMaass}
\mathfrak m_{n/2,j}=(g-1)(|n|-2j-1)\, .
\end{equation}

For each fixed even $n\neq0$, the associated graviton eigenvalue and total multiplicity are therefore
\begin{equation} \label{eq:MN1MaassGravMassMultiplicity}
\begin{aligned}
M^2L^2={}&3k(k+3)+\frac{3n^2}{4}-3\ell(\ell+1)
-3E^{\mathrm{univ}}_{n/2,j}\, ,\\[2pt]
\mathfrak d_{k,\ell,n;j}={}&(2\ell+1)(g-1)(|n|-2j-1)\, .
\end{aligned}
\end{equation}
For odd $n$, the Maass problem has half-integral weight and its global realisation depends on the spin structure induced by the fibration; we do not attempt to characterise this sector here. For nonzero even $n$, in addition to the universal values (\ref{eq:MN1MaassUniv}), there are generally further eigenvalues $E^{\Gamma}_{n/2,\alpha}$ that depend on the discrete group $\Gamma$ defining $\Sigma_2=H^2/\Gamma$.

The three sectors relevant here can consequently be summarised as follows, with the Maass line restricted to the explicitly characterised even-$n$, $n\neq0$, sector:
\begin{equation} \label{eq:MN1SpecSplit}
\begin{aligned}
{\cal S}_{\mathrm{put}}:\;& E_{n/2}=0\, ,\quad \varphi=1\ \text{locally}\, ,
\quad r=0,\ldots,k\, ,\quad n\in W_r\, ;\\[2pt]
{\cal S}_{\mathrm{glob}}^{(0)}:\;& E_0=0\, ,\quad \varphi=1\ \text{globally}\, ,
\quad n=0\, ,\quad r=0,2,\ldots,2\left\lfloor\frac{k}{2}\right\rfloor\, ;\\[2pt]
{\cal S}_{\mathrm{glob}}^{(\mathrm{Maass})}:\;&
E_{n/2}\in\big\{E^{\mathrm{univ}}_{n/2,j}\big\}
\cup\big\{E^{\Gamma}_{n/2,\alpha}\big\}\, ,
\quad \varphi\ \text{a global charged Maass eigensection}\, .
\end{aligned}
\end{equation}
The first line reproduces the local putative masses collected in appendix \ref{sec:PutSpec} by setting $E_{n/2}=0$ in (\ref{eq:QuantGravMass}). The second line is the U$(1)_0$-invariant, $\Sigma_2$-constant graviton sector retained in section~\ref{sec:GlobalSpecMN1}. The third line contains additional globally defined modes whose coefficients carry non-trivial dependence on $\Sigma_2$ and lie beyond the scope of this paper.

This discussion is the two-dimensional analogue of that in section 3.2 of \cite{Pico:2026rji}. There, the dependence of non-singlet modes on the compact hyperbolic three-manifold $\Sigma_3=H^3/\Gamma$ is governed by a Bochner connection Laplacian on vector bundles associated with representations of the generalised SO$(3)$ structure group. Here, the covariant Laplacian on the weight-$n/2$ automorphic line bundle is represented locally by the Maass operator (\ref{eq:MaassLap}), with the accompanying algebraic term already displayed in (\ref{eq:QuantGravMass}). In this sense, the weighted Maass problem is the rank-one U$(1)$ counterpart of the SO$(3)$ Bochner problem. In both cases, singlet harmonics admit globally defined constant coefficients and are insensitive to the twist, whereas non-singlets require solving the appropriate bundle-valued spectral problem. Accordingly, U$(1)_0$ invariance is a sufficient criterion for extracting a universal global sector from the putative spectrum, but it is not a necessary condition for the existence of a physical mode on a fixed compact quotient.

%%%%%%%%%%%%%%%
%%%%%%%%%%%%%%%

\section{Putative spectrum of MN1} \label{sec:PutSpec}

%%%%%%%%%%%%%%%
%%%%%%%%%%%%%%%

This appendix contains the putative KK spectrum of the MN1 solution. As defined in section~\ref{sec:PutUniv}, this is the spectrum that results from direct diagonalisation of the mass matrices of appendix \ref{sec:KKMassMat}, evaluated on the $D=5$ $\cN=8$ TCSO$(5,0,1;1)$ embedding tensor (\ref{eq:XSymbolsTilde}), the SO(5) generators (\ref{eq:TransCurlyT}) and the MN1 vacuum (\ref{eq:MN1vac}).

Define $r=2\ell$ as in (\ref{eq:MN1S3Eig}) and, from (\ref{eq:MN1DimFormula}),
{\setlength\arraycolsep{2pt}
\begin{eqnarray} \label{eq:MN1DimFormulaPartGen}
E^{(\textrm{grav})}_{k r n} &=&  E_{k \frac{r}{2} n \frac12\frac12} = 1 + \sqrt{4+3 k(k+3) +\tfrac34 n^2  -\tfrac34 r (r+2) } \; , \nonumber \\[4pt]
E^{(\textrm{gino})}_{k r n} &=&  E_{k \frac{r}{2} n \frac12 0} = 1 + \sqrt{\tfrac{11}{2} +3 k(k+3) +\tfrac34 n^2  -\tfrac34 r (r+2) } \; , \\[4pt]
E^{(\textrm{vec})}_{k r n} &=&  E_{k  \frac{r}{2} n 0 0} = 1 + \sqrt{7 +3 k(k+3) +\tfrac34 n^2  -\tfrac34 r (r+2) } \; . \nonumber
\end{eqnarray}
}We find that the putative spectrum of graviton multiplets contains a massless multiplet at KK level $k=0$, and a tower of long multiplets starting at $k \geq 1$: 
\begin{equation} \label{MN1GravGen}
A_1 \bar{A}_1 \big[ 3 ; \tfrac12 , \tfrac12 ; 0 \big] \otimes [0] \; \oplus \;
\bigoplus_{k=1}^{\infty} \bigoplus_{r=0}^{k}  \bigoplus_{n \in W_r}  L \bar{L} \big[ E^{(\textrm{grav})}_{krn}   ; \tfrac12 , \tfrac12 ; n \big] \otimes [\tfrac{r}{2}] \; .
\end{equation}
Here we have defined $W_r$ as in (\ref{eq:MN1WeightSet}). 
The spectrum of gravitino multiplets is, in turn, 
{\setlength\arraycolsep{0pt}
\begin{eqnarray} \label{MN1GinoGen}
&& \Big(
L\bar{A}_2
 \big[\tfrac72;\tfrac12,0;1\big]
\oplus
A_2\bar{L}
 \big[\tfrac72;0,\tfrac12;-1\big]
\Big)\otimes[0]
\nonumber \\[5pt]
&& \oplus\;
\Big(
L\bar{L}
 \big[1+\tfrac{\sqrt{13}}{2};0,\tfrac12;0\big]
\oplus
L\bar{L}
 \big[1+\tfrac{\sqrt{13}}{2};\tfrac12,0;0\big]
\Big)\otimes[\tfrac12]
\nonumber \\[5pt]
&& \oplus\;
\Big(
L\bar{L}
 \big[1+\tfrac{\sqrt{73}}{2};0,\tfrac12;1\big]
\oplus
L\bar{L}
 \big[1+\tfrac{\sqrt{73}}{2};\tfrac12,0;-1\big]
\Big)\otimes[0]
\nonumber \\[5pt]
&& \oplus\;
\Big(
L\bar{L}
 \big[1+\tfrac{\sqrt{73}}{2};0,\tfrac12;-1\big]
\oplus
L\bar{L}
 \big[1+\tfrac{\sqrt{73}}{2};\tfrac12,0;1\big]
\Big)\otimes[0]
\nonumber \\[5pt]
&& \oplus\;
\Big(
L\bar{L}
 \big[1+\tfrac{\sqrt{61}}{2};0,\tfrac12;0\big]
\oplus
L\bar{L}
 \big[1+\tfrac{\sqrt{61}}{2};\tfrac12,0;0\big]
\Big)\otimes[\tfrac12]
\nonumber \\[5pt]
&& \oplus\;
\Big(
L\bar{A}_1
 \big[\tfrac92;0,\tfrac12;1\big]
\oplus
A_1\bar{L}
 \big[\tfrac92;\tfrac12,0;-1\big]
\Big)\otimes[1]
\nonumber \\[5pt]
&& \oplus\;
\Big(
L\bar{L}
 \big[\tfrac92;0,\tfrac12;-1\big]
\oplus
L\bar{L}
 \big[\tfrac92;\tfrac12,0;1\big]
\Big)\otimes[1]
\\[5pt]
&& \oplus\;
\bigoplus_{k=2}^{\infty}
\bigoplus_{r=0}^{k+1}
\bigoplus_{s\in S_k(r)}
\bigoplus_{n\in W_s}
\Big(
L\bar{L}
 \big[E^{(\textrm{gino})}_{k rn};0,\tfrac12;n\big]
\oplus
L\bar{L}
 \big[E^{(\textrm{gino})}_{k rn};\tfrac12,0;n\big]
\Big)\otimes[\tfrac{r}{2}]
\; .
\nonumber
\end{eqnarray}
}The first two lines of (\ref{MN1GinoGen}) arise at KK level
$k=0$, the next five lines correspond to level $k=1$, and
the last line applies at $k\geq 2$. At $k=1$, the two
flavour-triplet pairs of dimension $\tfrac92$ consist of
one short pair and one strictly long pair with the
opposite charge--chirality assignment. The last line of
(\ref{MN1GinoGen}) uses the full signed weight set $W_s$
defined in (\ref{eq:MN1WeightSet}), with both Lorentz
chiralities occurring at each signed charge. The allowed
values of the auxiliary quantum number $s$ are
\begin{equation} \label{eq:Charges1Spec}
S_k(r) \equiv \{ s \in \mathbb{Z}_{\geq 0} \, | \, 0 \leq s \leq k , \,  |r-s| =1 \}  =
\left\{
\begin{array}{ll}
\{1 \} ,   &  r=0  \\[4pt]
\{ r-1 , r+1 \} ,   &  1 \leq r \leq k-1 \\[4pt]
\{ k-1 \} ,   &  r=k  \\[4pt]
\{ k \} ,   &  r=k+1  
 \end{array} \right. \; .
\end{equation}

\begin{table}[]

%\centering

\resizebox{\textwidth}{!}{

\begin{tabular}{l | l  } %\hline
\hline
\hline
\textbf{Multiplet} & $k=3$  \\[6pt]
\hline \\[-10pt]
Graviton
&
$L\bar{L}\left[1+\sqrt{58};\tfrac12,\tfrac12;0\right]\otimes[0]
\; \oplus \;
L\bar{L}\left[1+\sqrt{\tfrac{113}{2}};\tfrac12,\tfrac12;\pm1\right]
\otimes[\tfrac12] %$
%\\[5pt]
%%
%&
%$
\oplus\;
\Big(
L\bar{L}\left[1+\sqrt{55};\tfrac12,\tfrac12;\pm2\right]
\oplus
L\bar{L}\left[1+2\sqrt{13};\tfrac12,\tfrac12;0\right]
\Big)\otimes[1]$
\\[5pt]
&
$\oplus\;
\Big(
L\bar{L}\left[1+\sqrt{\tfrac{107}{2}};\tfrac12,\tfrac12;\pm3\right]
\oplus
L\bar{L}\left[1+\sqrt{\tfrac{95}{2}};\tfrac12,\tfrac12;\pm1\right]
\Big)\otimes[\tfrac32]$
\\[6pt]
\hline \\[-10pt]
Gravitino
&
$\bigoplus_{\rho=\pm1}\Big(
L\bar{L}\left[1+\tfrac{\sqrt{241}}{2};0,\tfrac12;\rho\right]
\oplus
L\bar{L}\left[1+\tfrac{\sqrt{241}}{2};\tfrac12,0;\rho\right]
\Big)\otimes[0]$
\\[5pt]
&
$\oplus\;
\Big(
\bigoplus_{\rho=\pm2}\Big(
L\bar{L}\left[1+\tfrac{\sqrt{241}}{2};0,\tfrac12;\rho\right]
\oplus
L\bar{L}\left[1+\tfrac{\sqrt{241}}{2};\tfrac12,0;\rho\right]\Big)
%$
%\\[-1pt]
%%
%&
%$\hspace{37mm}
%
\oplus
2L\bar{L}\left[1+\tfrac{\sqrt{229}}{2};0,\tfrac12;0\right]
\oplus
2L\bar{L}\left[1+\tfrac{\sqrt{229}}{2};\tfrac12,0;0\right]
\Big)\otimes[\tfrac12]$
\\[5pt]
&
$\oplus\;
\Big(
\bigoplus_{\rho=\pm3}\Big(
L\bar{L}\left[1+\tfrac{\sqrt{241}}{2};0,\tfrac12;\rho\right]
\oplus
L\bar{L}\left[1+\tfrac{\sqrt{241}}{2};\tfrac12,0;\rho\right]\Big)
%$
%\\[-1pt]
%%
%&
%$\hspace{37mm}
%
\oplus 
2\bigoplus_{\rho=\pm1}\Big(
L\bar{L}\left[1+\tfrac{\sqrt{217}}{2};0,\tfrac12;\rho\right]
\oplus
L\bar{L}\left[1+\tfrac{\sqrt{217}}{2};\tfrac12,0;\rho\right]\Big)
\Big)\otimes[1]$
\\[5pt]
&
$\oplus\;
\Big(
\bigoplus_{\rho=\pm2}\Big(
L\bar{L}\left[1+\tfrac{\sqrt{205}}{2};0,\tfrac12;\rho\right]
\oplus
L\bar{L}\left[1+\tfrac{\sqrt{205}}{2};\tfrac12,0;\rho\right]\Big)
%$
%\\[-1pt]
%%
%&
%$\hspace{37mm}
%
\oplus
L\bar{L}\left[1+\tfrac{\sqrt{193}}{2};0,\tfrac12;0\right]
\oplus
L\bar{L}\left[1+\tfrac{\sqrt{193}}{2};\tfrac12,0;0\right]
\Big)\otimes[\tfrac32]
$
\\[5pt]
&
$
\oplus\;
\Big(
\bigoplus_{\rho=\pm3}\Big(
L\bar{L}\left[1+\tfrac{\sqrt{193}}{2};0,\tfrac12;\rho\right]
\oplus
L\bar{L}\left[1+\tfrac{\sqrt{193}}{2};\tfrac12,0;\rho\right]\Big)
%$
%\\[-1pt]
%%
%&
%$\hspace{37mm}
%
\oplus
\bigoplus_{\rho=\pm1}\Big(
L\bar{L}\left[\tfrac{15}{2};0,\tfrac12;\rho\right]
\oplus
L\bar{L}\left[\tfrac{15}{2};\tfrac12,0;\rho\right]\Big)
\Big)\otimes[2]$
\\[6pt]
\hline \\[-10pt]
Vector
&
$\Big(
L\bar{L}\left[9;0,0;\pm2\right]
\oplus
2L\bar{L}\left[1+\sqrt{61};0,0;0\right]
\Big)\otimes[0] %$
%\\[5pt]
%%
%&
%$ 
%
\oplus
\Big(
L\bar{L}\left[1+\sqrt{\tfrac{131}{2}};0,0;\pm3\right]
\oplus
3L\bar{L}\left[1+\sqrt{\tfrac{119}{2}};0,0;\pm1\right]
\Big)\otimes[\tfrac12]$
\\[5pt]
&
$\oplus\;
\Big(
2L\bar{L}\left[1+\sqrt{58};0,0;\pm2\right]
\oplus
3L\bar{L}\left[1+\sqrt{55};0,0;0\right]
\Big)\otimes[1] %$
%\\[5pt]
%%
%&
%$
\oplus
\Big(
2L\bar{L}\left[1+\sqrt{\tfrac{113}{2}};0,0;\pm3\right]
\oplus
3L\bar{L}\left[1+\sqrt{\tfrac{101}{2}};0,0;\pm1\right]
\Big)\otimes[\tfrac32]$
\\[5pt]
&
$\oplus\;
\Big(
L\bar{L}\left[1+\sqrt{46};0,0;\pm2\right]
\oplus
L\bar{L}\left[1+\sqrt{43};0,0;0\right]
\Big)\otimes[2] %$
%\\[5pt]
%%
%&
%$
\oplus
\Big(
L\bar{L}\left[1+\sqrt{\tfrac{83}{2}};0,0;\pm3\right]
\oplus
L\bar{L}\left[1+\sqrt{\tfrac{71}{2}};0,0;\pm1\right]
\Big)\otimes[\tfrac52]$
\\[6pt]
\hline
\hline
& $k=2$  \\[6pt]
\hline \\[-10pt]
Graviton
&
$L\bar{L}\left[1+\sqrt{34};\tfrac12,\tfrac12;0\right]\otimes[0]
\; \oplus \;
L\bar{L}\left[1+\sqrt{\tfrac{65}{2}};\tfrac12,\tfrac12;\pm1\right]
\otimes[\tfrac12] %$
%\\[5pt]
%
%&
%$
\oplus 
\Big(
L\bar{L}\left[1+\sqrt{31};\tfrac12,\tfrac12;\pm2\right]
\oplus
L\bar{L}\left[1+2\sqrt{7};\tfrac12,\tfrac12;0\right]
\Big)\otimes[1]$
\\[6pt]
\hline \\[-10pt]
Gravitino
&
$\bigoplus_{\rho=\pm1}\Big(
L\bar{L}\left[1+\tfrac{\sqrt{145}}{2};0,\tfrac12;\rho\right]
\oplus
L\bar{L}\left[1+\tfrac{\sqrt{145}}{2};\tfrac12,0;\rho\right]
\Big)\otimes[0]$
\\[5pt]
&
$\oplus\;
\Big(
\bigoplus_{\rho=\pm2}\Big(
L\bar{L}\left[1+\tfrac{\sqrt{145}}{2};0,\tfrac12;\rho\right]
\oplus
L\bar{L}\left[1+\tfrac{\sqrt{145}}{2};\tfrac12,0;\rho\right]\Big)
%$
%\\[-1pt]
%%
%&
%$\hspace{37mm}
%
\oplus\;
2L\bar{L}\left[1+\tfrac{\sqrt{133}}{2};0,\tfrac12;0\right]
\oplus
2L\bar{L}\left[1+\tfrac{\sqrt{133}}{2};\tfrac12,0;0\right]
\Big)\otimes[\tfrac12]$
\\[5pt]
&
$\oplus\;
\bigoplus_{\rho=\pm1}\Big(
L\bar{L}\left[\tfrac{13}{2};0,\tfrac12;\rho\right]
\oplus
L\bar{L}\left[\tfrac{13}{2};\tfrac12,0;\rho\right]
\Big)\otimes[1]$
\\[5pt]
&
$\oplus\;
\Big(
\bigoplus_{\rho=\pm2}\Big(
L\bar{L}\left[1+\tfrac{\sqrt{109}}{2};0,\tfrac12;\rho\right]
\oplus
L\bar{L}\left[1+\tfrac{\sqrt{109}}{2};\tfrac12,0;\rho\right]\Big)
%$
%\\[-1pt]
%%
%&
%$\hspace{37mm}
%
\oplus
L\bar{L}\left[1+\tfrac{\sqrt{97}}{2};0,\tfrac12;0\right]
\oplus
L\bar{L}\left[1+\tfrac{\sqrt{97}}{2};\tfrac12,0;0\right]
\Big)\otimes[\tfrac32]$
\\[6pt]
\hline \\[-10pt]
Vector
&
$\Big(
L\bar{L}\left[1+2\sqrt{10};0,0;\pm2\right]
\oplus
2L\bar{L}\left[1+\sqrt{37};0,0;0\right]
\Big)\otimes[0]
%$
%\\[5pt]
%%
%&
%$
%
\oplus\;
2L\bar{L}\left[1+\sqrt{\tfrac{71}{2}};0,0;\pm1\right]
\otimes[\tfrac12]$
\\[5pt]
&
$\oplus\;
\Big(
2L\bar{L}\left[1+\sqrt{34};0,0;\pm2\right]
\oplus
3L\bar{L}\left[1+\sqrt{31};0,0;0\right]
\Big)\otimes[1]$
\\[5pt]
&
$\oplus\;
L\bar{L}\left[1+\sqrt{\tfrac{53}{2}};0,0;\pm1\right]
\otimes[\tfrac32] %$
%\\[5pt]
%%
%&
%$
%
\oplus\;
\Big(
L\bar{L}\left[1+\sqrt{22};0,0;\pm2\right]
\oplus
L\bar{L}\left[1+\sqrt{19};0,0;0\right]
\Big)\otimes[2]$
\\[6pt]
\hline
\hline
& $k=1$  \\[6pt]
\hline \\[-10pt]
Graviton
&
$L\bar{L} \left[5;\tfrac12,\tfrac12;0\right]\otimes[0]
\; \oplus \;
L\bar{L}\left[1+\sqrt{\tfrac{29}{2}};
\tfrac12,\tfrac12;\pm1\right]\otimes[\tfrac12]$
\\[5pt]
\hline \\[-10pt]
Gravitino
&
$\bigoplus_{\rho=\pm1}\Big(
L\bar{L}\left[1+\tfrac{\sqrt{73}}{2};0,\tfrac12;\rho\right]
\oplus
L\bar{L}\left[1+\tfrac{\sqrt{73}}{2};\tfrac12,0;\rho\right]
\Big)\otimes[0]
\; \oplus \;
\Big(
L\bar{L}\left[1+\tfrac{\sqrt{61}}{2};0,\tfrac12;0\right]
\oplus
L\bar{L}\left[1+\tfrac{\sqrt{61}}{2};\tfrac12,0;0\right]
\Big)\otimes[\tfrac12]$
\\[5pt]
&
$\oplus\;
\Big(
L\bar{A}_1\left[\tfrac92;0,\tfrac12;1\right]
\oplus
A_1\bar{L}\left[\tfrac92;\tfrac12,0;-1\right]
\Big)\otimes[1]$
\\[5pt]
&
$\oplus\;
\Big(
L\bar{L}\left[\tfrac92;0,\tfrac12;-1\right]
\oplus
L\bar{L}\left[\tfrac92;\tfrac12,0;1\right]
\Big)\otimes[1]$
\\[6pt]
\hline \\[-10pt]
Vector
&
$L\bar{L}\left[1+\sqrt{19};0,0;0\right]\otimes[0]
\; \oplus \;
2L\bar{L}\left[1+\sqrt{\tfrac{35}{2}};0,0;\pm1\right]
\otimes[\tfrac12]
\; \oplus \;
L\bar{L}\left[1+\sqrt{13};0,0;0\right]\otimes[1]
\; \oplus \;
L\bar{L}\left[1+\sqrt{\tfrac{17}{2}};0,0;\pm1\right]
\otimes[\tfrac32]$
\\[5pt]
&
$\oplus\;
\Big( L \bar{A}_2 \big[ 5 ; 0, 0  ; 2 \big] \oplus 
A_2 \bar{L} \big[ 5 ;  0 ,0  ; -2 \big] \Big) \otimes [1]$
\\[6pt]
\hline
\hline
& $k=0$  \\[6pt]
\hline \\[-10pt]
Graviton
&
$A_1\bar{A}_1\left[3;\tfrac12,\tfrac12;0\right]\otimes[0]$
\\[5pt]
\hline \\[-10pt]
Gravitino
&
$\Big(
L\bar{A}_2\left[\tfrac72;\tfrac12,0;1\right]
\oplus
A_2\bar{L}\left[\tfrac72;0,\tfrac12;-1\right]
\Big)\otimes[0]
\; \oplus \;
\Big(
L\bar{L}\left[1+\tfrac{\sqrt{13}}{2};\tfrac12,0;0\right]
\oplus
L\bar{L}\left[1+\tfrac{\sqrt{13}}{2};0,\tfrac12;0\right]
\Big)\otimes[\tfrac12]$
\\[6pt]
\hline \\[-10pt]
Vector
&
$A_2\bar{A}_2\left[2;0,0;0\right]\otimes[1]
\; \oplus \;
L\bar{L}\left[1+\sqrt{7};0,0;0\right]\otimes[0]$
\\[6pt]
\hline \\[-10pt]
Chiral
&
$ \Big( L\bar{B}_1\left[3;0,0;2\right] \oplus B_1\bar{L} \left[3;0,0;-2\right] \Big) \otimes[1]$
\\[6pt]
\hline
\hline
\end{tabular}

\qquad
}
\caption{\footnotesize{The first few levels $k$ of the putative spectrum of  $\mathrm{SU}(2,2|1)\times\mathrm{SU}(2)_{+}$
supermultiplets for MN1. Level $k=0$ reproduces the result of \cite{Varela:2025xeb}.
}\normalsize}
\label{tab:PutMN1Multiplets}
\end{table}

We also find the following vector multiplets:
{\setlength\arraycolsep{0pt}
\begin{eqnarray} \label{MN1VecGen}
&& A_2 \bar{A}_2 \big[ 2 ; 0 , 0 ; 0 \big] \otimes [1] \; \oplus \;
L \bar{L} \big[ 1+\sqrt{7}  ; 0 , 0 ; 0 \big] \otimes [0] %\nonumber \\[4pt]
 \nonumber \\[5pt]
&& \oplus \, \Big( L \bar{A}_2 \big[ 5 ; 0, 0  ; 2 \big] \oplus 
A_2 \bar{L} \big[ 5 ;  0 ,0  ; -2 \big] \Big) \otimes [1] \nonumber %
 \nonumber \\[3pt]
&&  \oplus \; \bigoplus_{k=1}^{\infty} \bigoplus_{s=0}^k \bigoplus_{n \in W_s}  \Big(  L \bar{L} \big[ E^{(\textrm{vec})}_{ksn} ; 0 , 0 ; n \big] \otimes [\tfrac{s}{2} ] \, \oplus \, 
\bigoplus_{r \in R_s} L \bar{L} \big[ E^{(\textrm{vec})}_{k rn} ;  0  , 0 ; n \big]  \  \otimes [\tfrac{r}{2}] \Big) \; .
\end{eqnarray}
}The first and second lines here respectively arise at $k=0$ and $k=1$, and contain the massless SU$(2)_+$ flavour current, a long singlet, and a pair of triplets of short vector multiplets of dimension $5$. The third line is generic for $k \geq 1$. It again utilises the range $W_s$ defined in (\ref{eq:MN1WeightSet}) along with 
\begin{equation} \label{eq:Charges2}
R_s  =
\left\{
\begin{array}{ll}
\{ 2 \} ,   &  s=0  \\[4pt]
\{ 1 , 3 \} ,   &  s=1 \\[4pt]
\{ s-2 , s , s+2  \} ,   &  s \geq 2 
 \end{array} \right. \; .
\end{equation}
Finally, there is a triplet of chiral and antichiral multiplets at KK level $k=0$,
\begin{equation} \label{MN1ChirGen}
\Big( L \bar{B}_1 \big[ 3 ; 0 , 0 ; 2 \big] 
\, \oplus \, 
B_1 \bar{L} \big[ 3 ; 0 , 0 ; -2 \big] \Big)  \otimes [1] \; ,
\end{equation}
which nevertheless do not extend at higher KK levels.

Except for the short multiplets, $A_1 \bar{A}_1$, $L \bar{A}_1$, $A_1 \bar{L}$, $A_2 \bar{A}_2$, $L \bar{A}_2$, $A_2 \bar{L}$, $L \bar{B}_1$, $B_1 \bar{L}$, explicitly listed in (\ref{MN1GravGen}), (\ref{MN1GinoGen}), (\ref{MN1VecGen}), (\ref{MN1ChirGen}), all multiplets in the spectrum are strictly long, without ever undergoing shortening. This is so even if some dimensions (\ref{eq:MN1DimFormulaPartGen}) become rational for certain values of the quantum numbers. In particular, for $k$ even, $E^{(\textrm{gino})}_{kk1} = \frac{1}{2} (3k+7)$ and, for $k$ odd, $E^{(\textrm{gino})}_{k,k+1,1} = \frac{3}{2} (k+2)$. At $k=1$, the single pair of short gravitino triplets $L \bar{A}_1$, $A_1 \bar{L}$ combines with the single pair of short vector triplets $L \bar{A}_2$, $A_2 \bar{L}$ into one pair of long gravitino triplets at threshold. The other pair of gravitino triplets at dimension $\tfrac92$, with the opposite charge--chirality assignment, is already strictly long and has no short-vector null pair. For convenience, table \ref{tab:PutMN1Multiplets} lists the first few KK levels of the putative MN1 spectrum.

\newpage

%%%%%%%%%%%%%%%
%%%%%%%%%%%%%%%

\bibliography{references}

%%%%%%%%%%%%%%%
%%%%%%%%%%%%%%%

\end{document}